\documentclass[useAMS,usenatbib]{mn2e}
\usepackage{graphicx}
\def \aj {AJ}
\def \mnras {MNRAS}

\def \apj {ApJ}
\def \apjs {ApJS}
\def \apss {Ap\&SS}

\def \aap {A\&A}
\def \nat {Nature}

\begin{document}

\title[The Type Ibn SN~2014av]{Massive stars exploding in a He-rich circumstellar medium - IX. SN~2014av, and characterization of Type Ibn SNe}
\author[Pastorello et al.]{A. Pastorello,$^1$\thanks{andrea.pastorello@oapd.inaf.it} X.-F. Wang,$^{2}$ F. Ciabattari,$^{3}$ D. Bersier,$^{4}$ P. A. Mazzali,$^{4}$ 
\newauthor X. Gao,$^{5}$  Z. Xu,$^{6}$  J.-J. Zhang,$^{7,8}$ S. Tokuoka,$^{9}$ S. Benetti,$^1$ E. Cappellaro,$^1$
\newauthor  N. Elias-Rosa,$^1$ A. Harutyunyan,$^{10}$ F. Huang,$^{2,11}$ M. Miluzio,$^{1,12}$ J. Mo,$^{2}$
\newauthor   P. Ochner,$^1$  L. Tartaglia,$^{1,13}$ G. Terreran,$^{1,14}$ L. Tomasella,$^1$ M. Turatto$^1$
\\
$^{1}$ INAF-Osservatorio Astronomico di Padova, Vicolo dell'Osservatorio 5, 35122 Padova, Italy \\
$^{2}$ Physics Department and Tsinghua Center for Astrophysics, Tsinghua University, Beijing, 100084, China\\
$^{3}$ Osservatorio Astronomico di Monte Agliale, Via Cune Motrone, 55023 Borgo a Mozzano, Lucca, Italy\\
$^{4}$ Astrophysics Research Institute, Liverpool John Moores University, IC2, Liverpool Science Park, 146 Brownlow Hill, Liverpool L3 5RF, UK \\
$^{5}$ Urumqi No.1 Senior High School, Urumqi, 830002, China\\
$^{6}$ Nanjing Putian Telecommunications Co., Nanjing, 210012, China\\ 
$^{7}$ Yunnan Observatories (YNAO), Chinese Academy of Sciences, Kunming 650011, China\\
$^{8}$ Key Laboratory for the Structure and Evolution of Celestial Objects, Chinese Academy of Sciences, Kunming 650011, China\\
$^{9}$ private address,  5-1-14 Tonda, Takatsuki, Osaka, 5690814, Japan\\
$^{10}$ Fundaci\'on Galileo Galilei-INAF, Telescopio Nazionale Galileo, Rambla Jos\'e Ana Fern\'andez P\'erez 7, 38712 Bre\~na Baja, TF,
 Spain\\
$^{11}$ Astronomical Department, Beijing Normal University, Beijing, 100875, China\\
$^{12}$ Instituto de Astrof\'isica de Canarias, C/ Vía L\'actea, s/n, E-38205 La Laguna, Santa Cruz de Tenerife, Spain\\
$^{13}$ Dipartimento di Fisica e Astronomia Galileo Galilei, Universit\`a di Padova, Vicolo dell'Osservatorio 3, Padova I-35122, Italy\\
$^{14}$ Astrophysics Research Centre, School of Mathematics and Physics, Queen's University Belfast, Belfast BT7 1NN, United Kingdom\\
}

\date{Accepted YYYY Month XX. Received YYYY Month XX;
  in original form YYYY Month XX}
\pagerange{\pageref{firstpage}--\pageref{lastpage}} \pubyear{YYYY}

\maketitle

\label{firstpage}

\begin{abstract} 
We present spectroscopic and photometric data of the Type Ibn supernova (SN) 2014av, discovered by 
the Xingming Observatory Sky Survey. Stringent pre-discovery detection limits indicate that the object 
was detected for the first time about 4 days after the explosion. A prompt follow-up campaign arranged by amateur astronomers 
allowed us to monitor the rising phase (lasting 10.6 days) and to accurately estimate the epoch of the maximum light, on 
2014 April 23 (JD = 2456771.1 $\pm$ 1.2). The absolute magnitude of the SN at the maximum light is 
$M_R = -19.76 \pm 0.16$. The post-peak light curve shows an initial fast decline
lasting about 3 weeks, and is followed by a slower decline in all bands until the end of the monitoring campaign.
The spectra are initially characterized by a hot continuum. Later on, the temperature declines and a number
of lines become prominent mostly in emission. In particular, later spectra are dominated by strong and narrow emission 
features of He I typical of Type Ibn supernovae (SNe), although there is a clear signature of lines from heavier elements
(in particular O I, Mg II and Ca II). A forest of relatively narrow Fe II lines
is also detected showing P-Cygni profiles, with the absorption component blue-shifted by about 1200 km s$^{-1}$. 
Another spectral feature often observed in interacting SNe, a strong blue pseudo-continuum, is seen 
in our latest spectra of SN~2014av.
We discuss in this paper the physical parameters of SN~2014av in the context of the Type Ibn supernova variety.

\end{abstract}


\maketitle

\section{Introduction} \label{intro}

Type Ibn SNe are a rare class of stripped-envelope events whose
spectra usually show relatively narrow lines of He I along with broader lines
of $\alpha$-elements similar to those observed in canonical Type Ib/c SNe \citep{pasto08a}.
The progenitors of SNe Ibn are thought to be hydrogen-poor Wolf-Rayet stars 
that have experienced major mass loss events shortly before the
terminal SN explosions. A pre-SN outburst was directly observed in the case of the prototypical SN~2006jc \citep{pasto07,fol07}.
The variety of observed properties characterizing this SN family has been discussed by \citet{tur14},
and several publications have been already devoted to this subject.\footnote{A comprehensive list of publications on Type Ibn SNe includes \citet{mat00},
\citet{seppo08}, \citet{smi08}, \citet{imm08}, \citet{elisa08}, \citet{noz08}, \citet{tom08}, \citet{anu09}, \citet{sak09}, \citet{chu09},
\citet{smi12}, \citet{san13}, \citet{mod14}, \citet{bia14}, \citet{gor14}, \citet{pasto08a,pasto08b,pasto14a,pasto14b,pasto14c}.}
However, although the number of Type Ibn SN discoveries is  growing, very
few  objects boast well-sampled data sets, and only occasionally they were
observed at early stages.

A rare opportunity to study a SN Ibn at very early phases has been provided
 with the discovery of a transient in the spiral galaxy UGC 4713, 
originally labelled as PSN J09002002+5229280, and later named with 
the IAU designation SN~2014av \citep{xu14}. 

\begin{figure}
\includegraphics[width=8.5cm,angle=0]{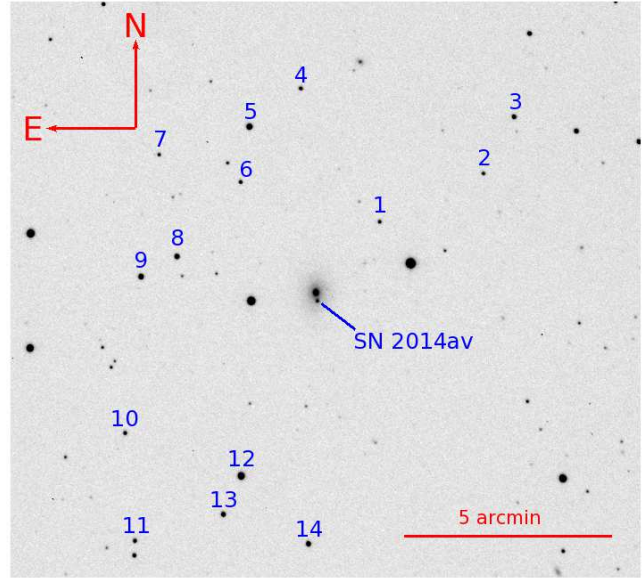}
\caption{Unfiltered amateur image of the field of UGC~4713. SN~2014av and the 14 local sequence stars used for the photometric calibration
are marked.
\label{fig1}}
\end{figure}

The new SN was discovered on 2014 April 19.72 UT (hereafter UT will be used along this paper), in the course of the 
Xingming Observatory Sky Survey (XOSS)\footnote{\it http://www.xjltp.com/xo/index-en.htm}, at a magnitude of about 16.2  \citep{xu14}.
 The earliest pre-discovery detection of the transient was registered on April 16.84 UT\footnote{Hereafter, UT times will be used throughout the paper.}
on images of the Italian Supernovae Search Project (ISSP)\footnote{\it http://italiansupernovae.org/}. 
Negative detections of the SN on images of UGC 4713 have been reported by the XOSS on March 26, 2014 and by the ISSP on April 6, 2014.
These amateur observations allowed us to constrain the explosion epoch with a very little uncertainty to April 13, 2014 (see Section \ref{ph}).
The new SN exploded about 2".7 West and 11".2 South of the core of the host galaxy (see Figure \ref{fig1}), which is
classified as Sb Type galaxy \citep[from {\it Hyperleda}\footnote{\it http://leda.univ-lyon1.fr/},][]{pat03}. 
UGC 4713 has a recessional velocity corrected for Local Group infall
into Virgo v$_{Vir}$ = 9225 $\pm$ 12 km s$^{-1}$ \citep[][from the NASA/IPAC Extragalactic Database; NED\footnote{\it https://ned.ipac.caltech.edu/}]{mou00}. 
Adopting a value for the Hubble Constant H$_0$ = 73 km s$^{-1}$ Mpc$^{-1}$ and
a standard cosmology ($\Omega_M$ = 0.27 and $\Omega_\lambda$ = 0.73),
we obtain a luminosity distance $d = 129 \pm 9$ Mpc (hence, a distance modulus $\mu = 35.56 \pm 0.15$ mag).
The Galactic extinction due to interstellar dust in the direction of SN~2014av is $A_B = 0.062$ mag \citep{sch11}, and there are
no spectroscopic signatures of additional reddening contribution in the host galaxy
(cfr. Section \ref{sp}).

\begin{table*}
\begin{center}
\caption{Landolt-calibrated magnitudes of the reference stars in the field of UGC 4713, and associated errors. \label{tab1} }
\begin{tabular}{cccccccc}
\hline \hline
Star ID & $R.A.$ & $Dec.$ & $U$ & $B$ & $V$ & $R$ & $I$ \\
 \hline
1  & 9:00:10.434 & 52:31:20.56 & 16.430  0.026 & 16.483  0.050 & 15.796  0.042 & 15.393  0.031 & 14.999  0.058 \\
2  & 8:59:54.384 & 52:32:30.42 & 17.332  0.028 & 16.902  0.054 & 16.066  0.043 & 15.588  0.033 & 15.161  0.061 \\
3  & 8:59:49.679 & 52:33:51.07 & 15.385  0.026 & 15.536  0.047 & 14.985  0.041 & 14.662  0.029 & 14.341  0.051 \\
4  & 9:00:22.854 & 52:34:28.95 & 16.052  0.019 & 16.164  0.038 & 15.583  0.028 & 15.245  0.028 & 14.917  0.052 \\
5  & 9:00:30.755 & 52:33:33.87 & 14.516  0.018 & 14.339  0.040 & 13.695  0.027 & 13.309  0.028 & 12.899  0.054 \\
6  & 9:00:32.061 & 52:32:15.37 & 16.441  0.020 & 16.419  0.041 & 15.742  0.028 & 15.350  0.030 & 14.983  0.055 \\
7  & 9:00:44.747 & 52:32:53.21 & 17.259  0.022 & 16.961  0.046 & 16.117  0.030 & 15.634  0.033 & 15.200  0.062 \\
8  & 9:00:41.823 & 52:30:28.80 & 15.266  0.029 & 15.037  0.043 & 14.386  0.033 & 14.008  0.031 & 13.646  0.051 \\
9  & 9:00:47.372 & 52:29:59.52 & 14.555  0.028 & 14.622  0.040 & 14.095  0.032 & 13.828  0.028 & 13.680  0.044 \\
10 & 9:00:49.557 & 52:26:17.28 & 17.943  0.033 & 16.941  0.059 & 15.805  0.038 & 15.142  0.045 & 14.518  0.086 \\
11 & 9:00:47.894 & 52:23:44.54 & 16.140  0.029 & 16.051  0.041 & 15.465  0.032 & 15.114  0.031 & 14.745  0.052 \\
12 & 9:00:31.525 & 52:25:17.66 & 13.652  0.028 & 13.491  0.032 & 12.916  0.015 & 12.578  0.025 & 12.240  0.047 \\
13 & 9:00:34.258 & 52:24:22.86 & 17.542  0.031 & 16.393  0.066 & 15.092  0.041 & 14.328  0.052 & 13.607  0.101 \\
14 & 9:00:21.035 & 52:23:42.40 & 15.557  0.029 & 15.328  0.044 & 14.614  0.033 & 14.185  0.035 & 13.737  0.061 \\ \hline
\end{tabular}
\end{center}
\end{table*}

\citet{zan14}, on the basis of the blue spectral continuum and a presence of 
relatively narrow He I lines, classified SN~2014av as a young Type Ibn SN similar
to SN~2006jc. The young age of the SN and the rarity of Type Ibn events \citep[$\le$ 2 per cent of core-collapse SNe,][]{pasto14a}
motivated our team to initiate a monitoring campaign of this new member of the family.
Photometric and spectroscopic observations of SN~2014av are presented in Sections \ref{ph} and \ref{sp}, respectively;
a bolometric light curve and an estimate of the explosion parameters are given in Section \ref{bolom}. 
An extensive characterization of spectral properties of SNe Ibn is provided in Section \ref{sp_cha}. The properties of the progeniors of SNe Ibn
are illustrated in Section \ref{prog} and a summary follows in Section \ref{summary}.

\section{Photometric Observations} \label{ph}

Our multi-band (Johnson-Cousins $UBVRI$ and Sloan $griz$) photometric follow-up campaign started when the object had already reached the 
maximum light, and continued up to phase $\sim$70 days after discovery, until the object 
became unobservable because it was  in solar conjunction.
The instruments used in our photometric campaign are the 3.58-m Telescopio Nazionale Galileo (TNG) equipped with Dolores,
the 2.56-m Nordic Optical Telescope (NOT) with ALFOSC and the 2.0-m Liverpool Telescope with IO:O, all located at Roque de los 
Muchachos Observatory, La Palma, Canary Islands (Spain); the 1.82-m Copernico 
Telescope of Mt. Ekar, Asiago (Italy) equipped with AFOSC.
In order to sample the early-time SN evolution, we included in our dataset unfiltered observations
from amateur astronomers. The instruments used are listed as a footnote in the SN photometry tables.
These additional data, essential to constrain the rising branch of the light curve and the peak luminosity, 
were scaled to $V$ or $R$-band (and similarly to Sloan $g$ {\bf or} $r$-band) photometry, depending on the approximate wavelength
 of the peak of the quantum efficiency curves of the CCDs used in the observations.

\begin{figure*}
{\includegraphics[width=8.72cm,angle=0]{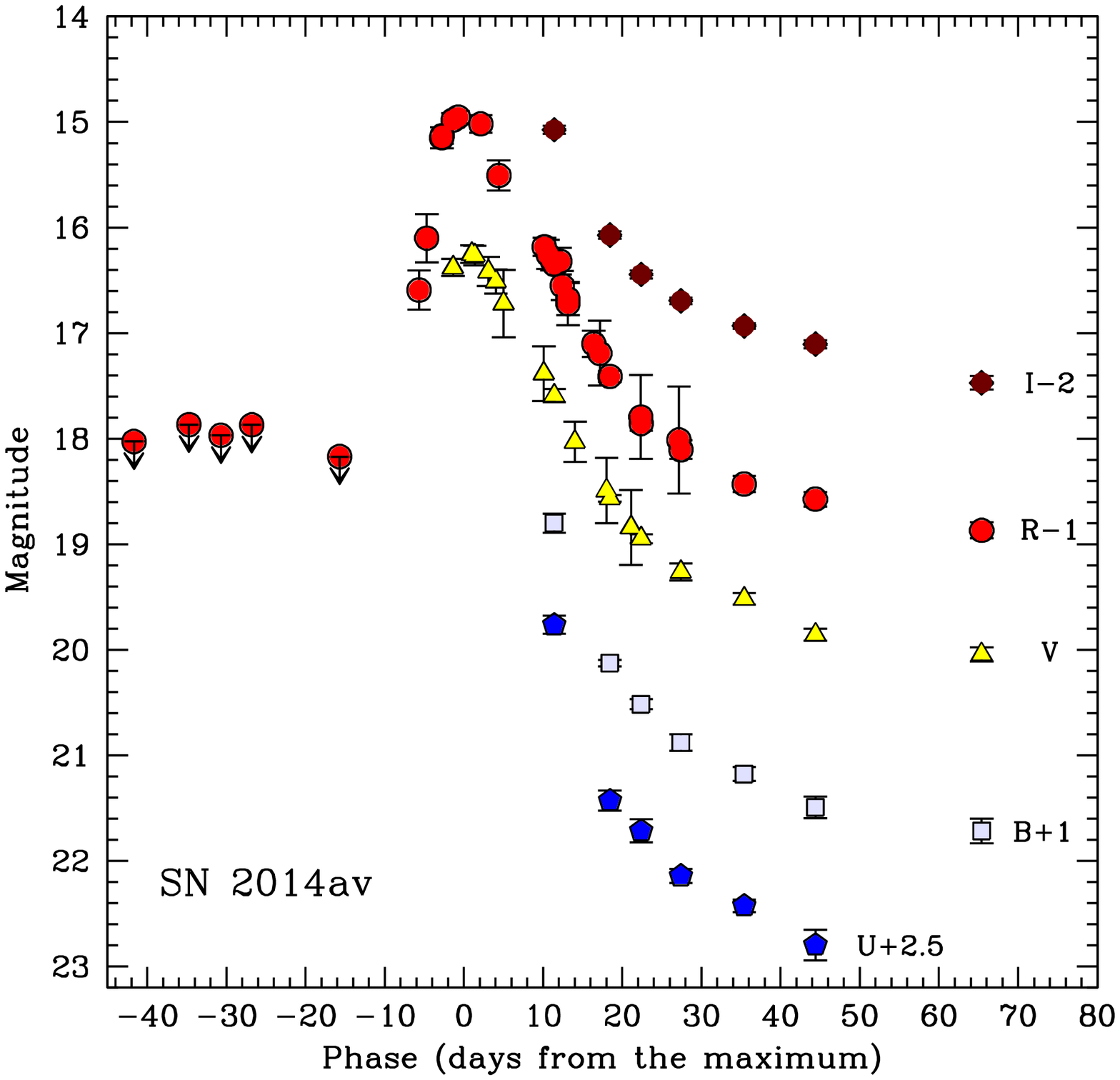}
\includegraphics[width=8.72cm,angle=0]{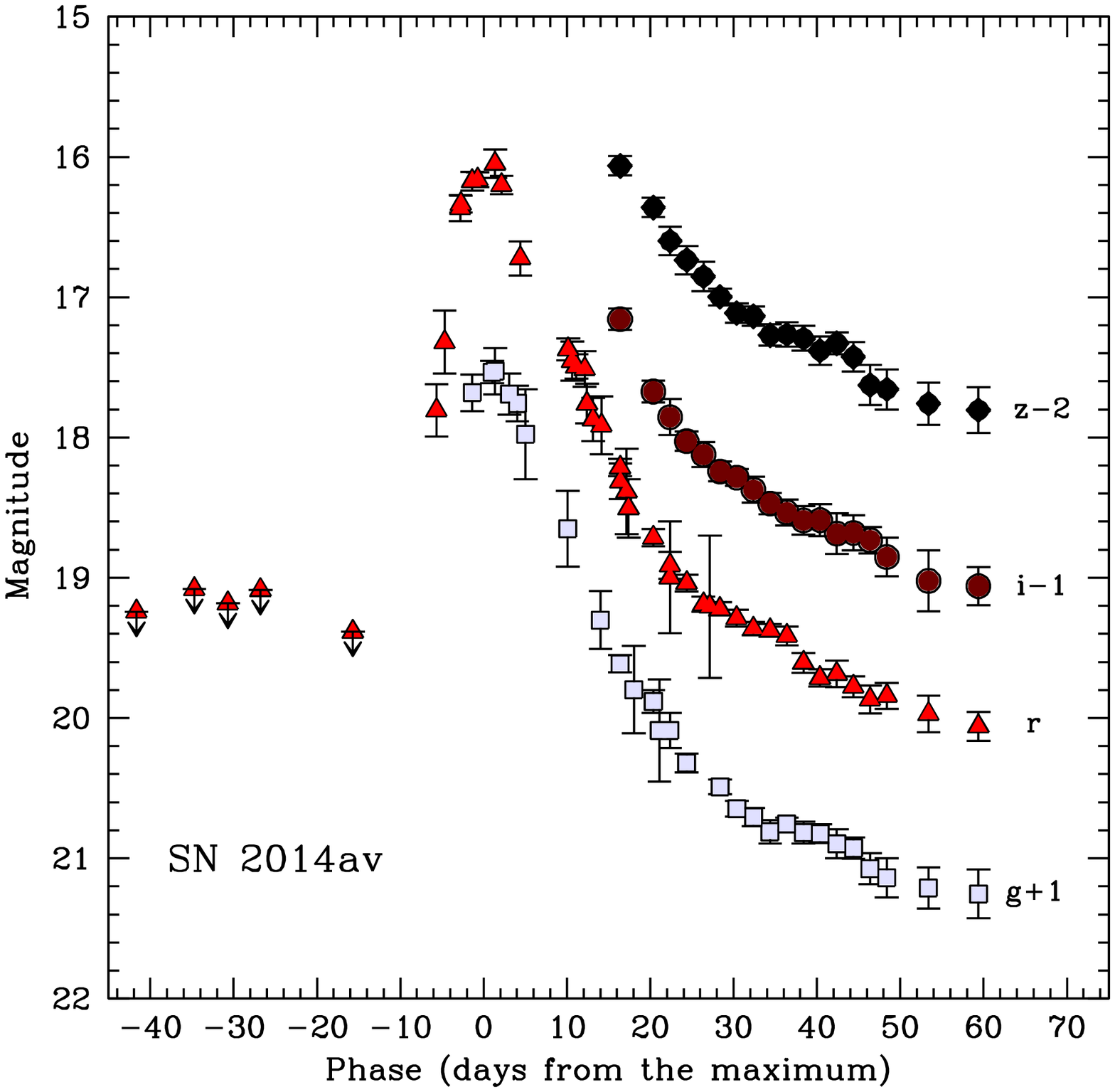}
\caption{$UBVRI$ (left) and $griz$ (right) light curves of SN~2014av. The closest pre-discovery limits are also shown.
Data from unfiltered images are rescaled to Johnson-Cousins $V$ or $R$ (vegamag), and Sloan $g$ or $r$ (ABmag), depending on the 
wavelength of the peak of the quantum efficiency curves of the CCDs used for the observations, as specified in the text.}
\label{fig2}}
\end{figure*}

All frames were pre-processed using standard procedures in \textsc{IRAF}. We corrected the frames for 
bias, overscan and flat-fielding, and they were finally astrometrically calibrated.   
For photometric measurements, we used the dedicated pipeline \textsc{SNOoPY} \citep{cap14}. 
This pipeline consists of a collection of \textsc{PYTHON} scripts calling standard \textsc{IRAF} tasks, and other data
analysis tools such as \textsc{SEXTRACTOR} for source extraction and classification, \textsc{DAOPHOT}
for point spread function (PSF) fitting, and \textsc{HOTPANTS}\footnote{\it http://www.astro.washington.edu/users/becker/v2.0/hotpants.html}
for PSF matching and image subtraction. PSF-fitting photometry provided reliable results in the high-quality,
high signal-to-noise (S/N) filtered images from professional telescopes, whilst this measurement method provided
inaccurate results in lower S/N unfiltered images from amateur astronomers. For the latter epochs, we employed the
template subtraction method to remove the contaminating flux from the host galaxy background. Good-quality pre-explosion images of
UGC 4713 were used as templates. These images were available after the routine monitoring observations of UGC~4713
obtained in the course of the XOSS and ISSP  SN searches.

\begin{table*}
\begin{center}
\caption{Johnson-Cousins magnitudes of SN~2014av, and associated errors. \label{tab2} }
\begin{tabular}{cccccccc}
\hline \hline
Obs. date   & Average $JD$ & $U$ & $B$ & $V$ & $R$ & $I$ & Instrument \\
 \hline
2010/03/14  &    2455270.39 &      --      &      --      &       --      &  $>$19.26     &      --      &  1 \\ 
2011/03/09  &    2455630.37 &      --      &      --      &       --      &  $>$19.27     &      --      &  1 \\  
2013/12/03  &    2456630.49 &      --      &      --      &       --      &  $>$18.93     &      --      &  1 \\  
2013/12/10  &    2456637.58 &      --      &      --      &       --      &  $>$19.19     &      --      &  1 \\  
2013/12/19  &    2456645.55 &      --      &      --      &       --      &  $>$19.17     &      --      &  1 \\  
2014/01/06  &    2456664.49 &      --      &      --      &       --      &  $>$18.52     &      --      &  1 \\  
2014/03/11  &    2456728.33 &      --      &      --      &       --      &  $>$19.03     &      --      &  1 \\  
2014/03/18  &    2456735.29 &      --      &      --      &       --      &  $>$18.87     &      --      &  1 \\  
2014/03/22  &    2456739.32 &      --      &      --      &       --      &  $>$18.97     &      --      &  2 \\  
2014/03/26  &    2456743.23 &      --      &      --      &       --      &  $>$18.87     &      --      &  2 \\  
2014/04/06  &	 2456754.31 &      --      &      --      &       --      &  $>$19.17     &      --      &  1 \\  
2014/04/16  &	 2456764.34 &      --      &      --      &       --      &  17.592 0.185 &      --      &   1 \\ 
2014/04/17  &	 2456765.32 &      --      &      --      &       --      &  17.100 0.229 &      --      &   1 \\ 
2014/04/19  &    2456767.22 &      --      &      --      &       --      &  16.150 0.098 &      --      &   2 \\  
2014/04/19  &    2456767.27 &      --      &      --      &       --      &  16.130 0.079 &      --      &   2 \\  
2014/04/21  &    2456768.63 &      --      &      --      &  16.379 0.080 &  15.983 0.067 &      --      &   3 \\  
2014/04/21  &    2456769.28 &      --      &      --      &       --      &  15.954 0.058 &      --      &   4 \\  
2014/04/23  &    2456771.03 &      --      &      --      &  16.253 0.083 &       --      &      --      &   5 \\  
2014/04/23  &    2456771.38 &      --      &      --      &  16.266 0.094 &       --      &      --      &   6 \\  
2014/04/24  &    2456772.14 &      --      &      --      &       --      &  16.018 0.083 &      --      &   2 \\  
2014/04/25  &    2456773.09 &      --      &      --      &  16.417 0.138 &       --      &      --      &   5 \\  
2014/04/26  &    2456774.04 &      --      &      --      &  16.512 0.114 &       --      &      --      &   5 \\  
2014/04/26  &	 2456774.41 &      --      &      --      &       --      &  16.507 0.143 &      --      &   1 \\  
2014/04/27  &    2456775.02 &      --      &      --      &  16.720 0.320 &       --      &      --      &   5 \\  
2014/05/02  &    2456780.10 &      --      &      --      &  17.383 0.260 &       --      &      --      &   5 \\  
2014/05/02  &    2456780.16 &      --      &      --      &       --      &  17.182 0.087 &      --      &   2 \\  
2014/05/03  &    2456780.63 &      --      &      --      &       --      &  17.254 0.139 &      --      &   7 \\  
2014/05/03  &    2456781.14 &      --      &      --      &       --      &  17.311 0.086 &      --      &   2 \\  
2014/05/03  &    2456781.41 & 17.263 0.087 & 17.800 0.090 &  17.592 0.063 &  17.348 0.033 & 17.073 0.037 &   8 \\ 
2014/05/04  &    2456782.13 &      --      &      --      &       --      &  17.320 0.126 &      --      &   2 \\  
2014/05/04  &    2456782.42 &      --      &      --      &       --      &  17.549 0.140 &      --      &   9 \\  
2014/05/05  &    2456783.14 &      --      &      --      &       --      &  17.676 0.153 &      --      &   2 \\  
2014/05/06  &    2456784.03 &      --      &      --      &  18.031 0.191 &       --      &      --      &   5 \\  
2014/05/06  &    2456784.15 &      --      &      --      &       --      &  17.720 0.204 &      --      &   2 \\  
2014/05/08  &    2456786.40 &      --      &      --      &       --      &  18.101 0.125 &      --      &   1 \\  
2014/05/09  &    2456787.16 &      --      &      --      &       --      &  18.191 0.307 &      --      &   2 \\  
2014/05/10  &    2456788.02 &      --      &      --      &  18.491 0.309 &       --      &      --      &   5 \\  
2014/05/10  &	 2456788.45 & 18.928 0.097 & 19.125 0.030 &  18.566 0.031 &  18.410 0.051 & 18.071 0.034 &  10 \\  
2014/05/13  &    2456791.10 &      --      &      --      &  18.841 0.354 &       --      &      --      &   5 \\  
2014/05/14  &    2456792.36 &      --      &      --      &       --      &  18.794 0.397 &      --      &   1 \\  
2014/05/14  &	 2456792.39 & 19.215 0.109 & 19.515 0.047 &  18.946 0.043 &  18.856 0.069 & 18.444 0.039 &   8 \\ 
2014/05/19  &    2456797.16 &      --      &      --      &       --      &  19.012 0.507 &      --      &   2 \\  
2014/05/19  &    2456797.41 & 19.641 0.067 & 19.877 0.078 &  19.263 0.081 &  19.104 0.089 & 18.694 0.030 &   8 \\  
2014/05/27  &	 2456805.40 & 19.924 0.059 & 20.177 0.066 &  19.518 0.056 &  19.430 0.075 & 18.931 0.027 &   8 \\  
2014/06/05  &    2456814.39 & 20.296 0.145 & 20.492 0.101 &  19.857 0.056 &  19.573 0.068 & 19.105 0.038 &   8 \\ 
2014/06/26  &	 2456835.39 &      --      & 20.715 0.116 &  20.043 0.065 &  19.869 0.076 & 19.471 0.064 &   8 \\  \hline
\end{tabular}
\\
\begin{flushleft}
1 = 0.50-m Newton telescope (Lotti) + FLI Proline 4710 CCD (ISSP collaboration, Osservatorio Astronomico di Monte Agliale, Borgo a Mozzano, Lucca, Italy); 
2 = 0.36-m Celestron C14 telescope +  QHY9 + KAF8300 CCD  (Obs. Z. Xu $\&$ X. Gao, Xingming Observatory; Xiaofeng, Gangou, China);
3 = 0.36-m Celestron C14 telescope + DSI Pro-DSI ll CCD (Obs. W. S. Wiethoff; SOLO Observatory, Port Wing, Wisconsin, USA);
4 = 0.43-m Planewave Corrected Dall-Kirkham Astrograph + Paramount ME + SBIG STL-6303E CCD (Obs. G. Masi, The Virtual Telescope Project 2.0, 
Bellatrix Astronomical Observatory, Ceccano, Frosinone, Italy);
5 = 0.25-m Whitey Dob telescope +  Starlight Xpress SXVR-H694 CCD (Obs. S. Tokuoka, Takatsuki, Osaka, Japan);
6 = 0.25-m Meade SC telescope + Atik 314L+ Sony ICX-285AL CCD (Obs. G. Locatelli; Maritime Alps Observatory MPC K32, Cuneo, Italy); 
7 = 0.28-m Celestron C11 telescope + Orion StarShoot DSMI III CCD (Obs. S. Howerton; Arkansas City, Kansas, USA); 
8 = 3.58-m TNG + Dolores (Roque de los Muchachos, La Palma, Canary Islands, Spain); 
9 = 0.28-m Celestron C11 telescope + SBIG STT-1603 ME camera (Obs. J.-M. Llapasset; Perpignan Observatory, France); 
10 = 2.56-m NOT + ALFOSC (Roque de los Muchachos, La Palma, Canary Islands, Spain).
\end{flushleft}

\end{center}
\end{table*}

\begin{table*}
\begin{center}
\caption{Sloan magnitudes of SN~2014av, and associated errors. \label{tab3} }
\begin{tabular}{ccccccc}
\hline \hline
Obs. date   & Average $JD$ & $g$ & $r$ &$ i$ & $z$ & Instrument \\
 \hline
2010/03/14 & 55270.390  &      -- &  $>$19.48   &      -- &       -- &  1 \\
2011/03/09 & 55630.374  &      -- &  $>$19.47   &      -- &       -- &  1 \\
2013/12/03 & 56630.490  &      -- &  $>$19.15   &      -- &       -- &  1 \\
2013/12/10 & 56637.581  &      -- &  $>$19.38   &      -- &       -- &  1 \\
2013/12/19 & 56645.553  &      -- &  $>$19.39   &      -- &       -- &  1 \\
2014/01/06 & 56664.491  &      -- &  $>$18.70   &      -- &       -- &  1 \\
2014/03/11 & 56728.331  &      -- &  $>$19.24   &      -- &       -- &  1 \\ 
2014/03/18 & 56735.287  &      -- &  $>$19.08   &      -- &       -- &  1 \\
2014/03/22 & 56739.318  &      -- &  $>$19.18   &      -- &       -- &  2 \\
2014/03/26 & 56743.226  &      -- &  $>$19.09   &      -- &       -- &  2 \\
2014/04/06 & 56754.313  &      -- &  $>$19.39   &      -- &       -- &  1 \\
2014/04/16 & 56764.341  &      -- &  17.808 0.187  &      -- &       -- &   1 \\
2014/04/17 & 56765.316  &      -- &  17.321 0.224  &      -- &       -- &   1 \\
2014/04/19 & 56767.217  &      -- &  16.365 0.092  &      -- &       -- &   2 \\
2014/04/19 & 56767.267  &      -- &  16.337 0.059  &      -- &       -- &   2 \\
2014/04/21 & 56768.632  & 16.681 0.131 &  16.174 0.066  &      -- &       -- & 3 \\
2014/04/21 & 56769.284  &      -- &  16.162 0.053  &      -- &       -- &   4 \\
2014/04/23 & 56771.030  & 16.534 0.080 &       --  &      -- &       -- &  5 \\
2014/04/23 & 56771.381  & 16.529 0.165 &  16.049 0.100  &      -- &       -- & 6 \\
2014/04/24 & 56772.138  &      -- &  16.201 0.065  &      -- &       -- &   2 \\
2014/04/25 & 56773.089  & 16.690 0.147 &       --  &      -- &       -- &  5 \\
2014/04/26 & 56774.042  & 16.758 0.125 &       --  &      -- &       -- &  5 \\
2014/04/26 & 56774.413  &      -- &  16.723 0.121  &      -- &       -- &   1 \\
2014/04/27 & 56775.024  & 16.978 0.320 &       --  &      -- &       -- &  5 \\
2014/05/02 & 56780.099  & 17.652 0.270 &       --  &      -- &       -- &  5 \\
2014/05/02 & 56780.158  &      -- &  17.373 0.077  &      -- &       -- &   2 \\
2014/05/03 & 56780.632  &      -- &  17.455 0.140  &      -- &       -- &  7 \\
2014/05/03 & 56781.136  &      -- &  17.495 0.087  &      -- &       -- &   2 \\
2014/05/04 & 56782.131  &      -- &  17.512 0.127  &      -- &       -- &   2 \\
2014/05/04 & 56782.417  &      -- &  17.759 0.141  &      -- &       -- &  8 \\
2014/05/05 & 56783.143  &      -- &  17.872 0.155  &      -- &       -- &   2 \\
2014/05/06 & 56784.031  & 18.301 0.205 &       --  &      -- &       -- &  5 \\
2014/05/06 & 56784.153  &      -- &  17.915 0.206  &      -- &       -- &   2 \\
2014/05/08 & 56786.385  & 18.613 0.061 &  18.215 0.062  & 18.157 0.075 &  18.064 0.068 &  9 \\
2014/05/08 & 56786.404  &      -- &  18.313 0.127  &      -- &       -- &   1 \\
2014/05/09 & 56787.158  &      -- &  18.385 0.305  &      -- &       -- &   2 \\
2014/05/09 & 56787.407  &      -- &  18.505 0.208  &      -- &       -- &   10 \\
2014/05/10 & 56788.023  & 18.797 0.312 &       --  &      -- &       -- &  5 \\
2014/05/12 & 56790.370  & 18.883 0.082 &  18.714 0.062  & 18.673 0.078 &  18.361 0.068 &   9 \\ 
2014/05/13 & 56791.096  & 19.089 0.363 &       --  &      -- &       -- &  5 \\
2014/05/14 & 56792.361  &      -- &  18.998 0.398  &      -- &       -- &   1 \\
2014/05/14 & 56792.370  & 19.088 0.124 &  18.911 0.095  & 18.855 0.128 &  18.599 0.103 &  9 \\  
2014/05/16 & 56794.370  & 19.321 0.067 &  19.038 0.059  & 19.027 0.068 &  18.736 0.101 &  9 \\  
2014/05/18 & 56796.385  &      --      &  19.189 0.053  & 19.122 0.090 &  18.852 0.106 &  9 \\  
2014/05/09 & 56797.159  &      --      &  19.206 0.506  &      --      &       --      &  2 \\
2014/05/20 & 56798.380  & 19.490 0.053 &  19.222 0.048  & 19.242 0.070 &  18.998 0.060 &  9 \\  
2014/05/22 & 56800.375  & 19.646 0.056 &  19.288 0.061  & 19.286 0.060 &  19.114 0.068 &  9 \\  
2014/05/24 & 56802.380  & 19.706 0.065 &  19.363 0.048  & 19.373 0.092 &  19.136 0.068 &  9 \\  
2014/05/26 & 56804.380  & 19.811 0.084 &  19.376 0.046  & 19.470 0.076 &  19.270 0.076 &  9 \\  
2014/05/28 & 56806.395  & 19.754 0.045 &  19.414 0.067  & 19.534 0.092 &  19.265 0.087 &  9 \\  
2014/05/30 & 56808.385  & 19.817 0.078 &  19.606 0.070  & 19.591 0.100 &  19.294 0.089 &  9 \\  
2014/06/01 & 56810.385  & 19.823 0.066 &  19.715 0.061  & 19.589 0.114 &  19.381 0.101 &  9 \\  
2014/06/03 & 56812.390  & 19.896 0.103 &  19.685 0.093  & 19.686 0.144 &  19.328 0.077 &  9 \\ 
2014/06/05 & 56814.385  & 19.927 0.075 &  19.777 0.073  & 19.680 0.126 &  19.426 0.105 &  9 \\  
2014/06/07 & 56816.385  & 20.074 0.110 &  19.867 0.099  & 19.732 0.094 &  19.626 0.144 &  9 \\  
2014/06/09 & 56818.390  & 20.138 0.139 &  19.843 0.092  & 19.852 0.138 &  19.657 0.143 &  9 \\  
2014/06/14 & 56823.395  & 20.212 0.147 &  19.971 0.131  & 20.022 0.216 &  19.759 0.150 &  9 \\  
2014/06/20 & 56829.390  & 20.253 0.173 &  20.059 0.104  & 20.060 0.137 &  19.805 0.162 &  9 \\  \hline
\end{tabular}
\\
\begin{flushleft}
1 = 0.51-m Lotti telescope + FLI Proline 4710 CCD (ISSP collaboration, Osservatorio di Monte Agliale, Borgo a Mozzano, Lucca, Italy); 
2 = 0.36-m Celestron C14 telescope +  QHY9 + KAF8300 CCD (Obs. Z. Xu $\&$ X. Gao, Xingming Observatory; Xiaofeng, Gangou, China);
3 = 0.36-m Celestron C14 telescope + DSI Pro-DSI ll CCD (Obs. W. S. Wiethoff; SOLO Observatory, Port Wing, Wisconsin, USA);
4 = 0.43-m Planewave Corrected Dall-Kirkham Astrograph + Paramount ME + SBIG STL-6303E CCD (Obs. G. Masi, The Virtual Telescope Project 2.0, 
Bellatrix Astronomical Observatory, Ceccano, Frosinone, Italy);
5 = 0.25-m Whitey Dob telescope +  Starlight Xpress SXVR-H694 CCD (Obs. S. Tokuoka, Takatsuki, Osaka, Japan);
6 = 0.25-m Meade SC telescope + Atik 314L + Sony ICX-285AL CCD (Obs. G. Locatelli; Maritime Alps Observatory MPC K32, Cuneo, Italy); 
7 = 0.28-m Celestron C11 telescope + Orion StarShoot DSMI III CCD (Obs. S. Howerton; Arkansas City, Kansas, USA); 
8 = 0.28-m Celestron C11 telescope + SBIG STT-1603 ME camera (Obs. J.-M. Llapasset; Perpignan Observatory, France); 
9 = 2.0-m Liverpool Telescope + IO:O camera (Roque de los Muchachos, La Palma, Canary Islands, Spain);
10 = 1.82-m Copernico Telescope (Mt. Ekar, Asiago Observatory, Italy). 
\end{flushleft}

\end{center}
\end{table*}

\begin{table}
\begin{center}
\caption{Main light curve parameters for SN~2014av. \label{tab4} }
\begin{tabular}{cccc}
\hline \hline
Filter   & peak magnitude & $\gamma_{0-25}^\ddag$ & $\gamma_{25-60}^\ddag$ \\ \hline
$U$ & --             & 18.47 $\pm$ 4.24 & 3.86 $\pm$ 0.17 \\
$B$ & --             & 16.00 $\pm$ 2.32 & 2.12 $\pm$ 0.49 \\
$V$ & 16.24$\pm$0.07 & 13.40 $\pm$ 0.26 & 2.00 $\pm$ 0.50 \\
$R$ & 15.85$\pm$0.05 & 12.73 $\pm$ 0.55 & 1.87 $\pm$ 0.36 \\
$I$ & --             & 12.69 $\pm$ 1.22 & 1.98 $\pm$ 0.17 \\ 
$g$ & 16.52$\pm$0.04 & 13.11 $\pm$ 0.46 & 2.63 $\pm$ 0.21 \\
$r$ & 16.07$\pm$0.05 & 12.98 $\pm$ 0.46 & 3.07 $\pm$ 0.12 \\
$i$ & --             & 11.85 $\pm$ 0.95 & 2.84 $\pm$ 0.17 \\
$z$ & --             & 8.61 $\pm$ 0.58 & 2.93 $\pm$ 0.22 \\
 \hline
\end{tabular}
\begin{flushleft}

$^\ddag$ in mag/100$^d$ units
\end{flushleft}
\end{center}
\end{table}

Once instrumental magnitudes of the SN and a number of stellar sources in the SN vicinity were obtained, 
we accurately calibrated the Johnson-Cousins magnitudes of a sequence of local standards in the field. To
this aim, we selected observations obtained during a few photometric nights in which standard photometric 
fields from the catalogue of \citet{lan92} were also observed. These observations allowed us to obtain 
the zero-points and colour terms for each night and each specific instrumental set-up. 
The resulting calibrated magnitudes of the sequence stars in the SN field were derived by averaging 
the measures obtained during these selected nights (see Figure \ref{fig1} and Table \ref{tab1}). This allowed us to correct
the zero-points estimated in non photometric nights and accurately calibrate the SN magnitudes.
Sloan-band photometry was directly calibrated using the SDSS DR10 catalogue. 
The final Johnson-Cousins and Sloan magnitudes of the SN are reported in Table \ref{tab2} and Table \ref{tab3},
respectively, and the resulting light curves are shown in Figure \ref{fig2}.

\begin{figure*}
\includegraphics[width=13.0cm,angle=270]{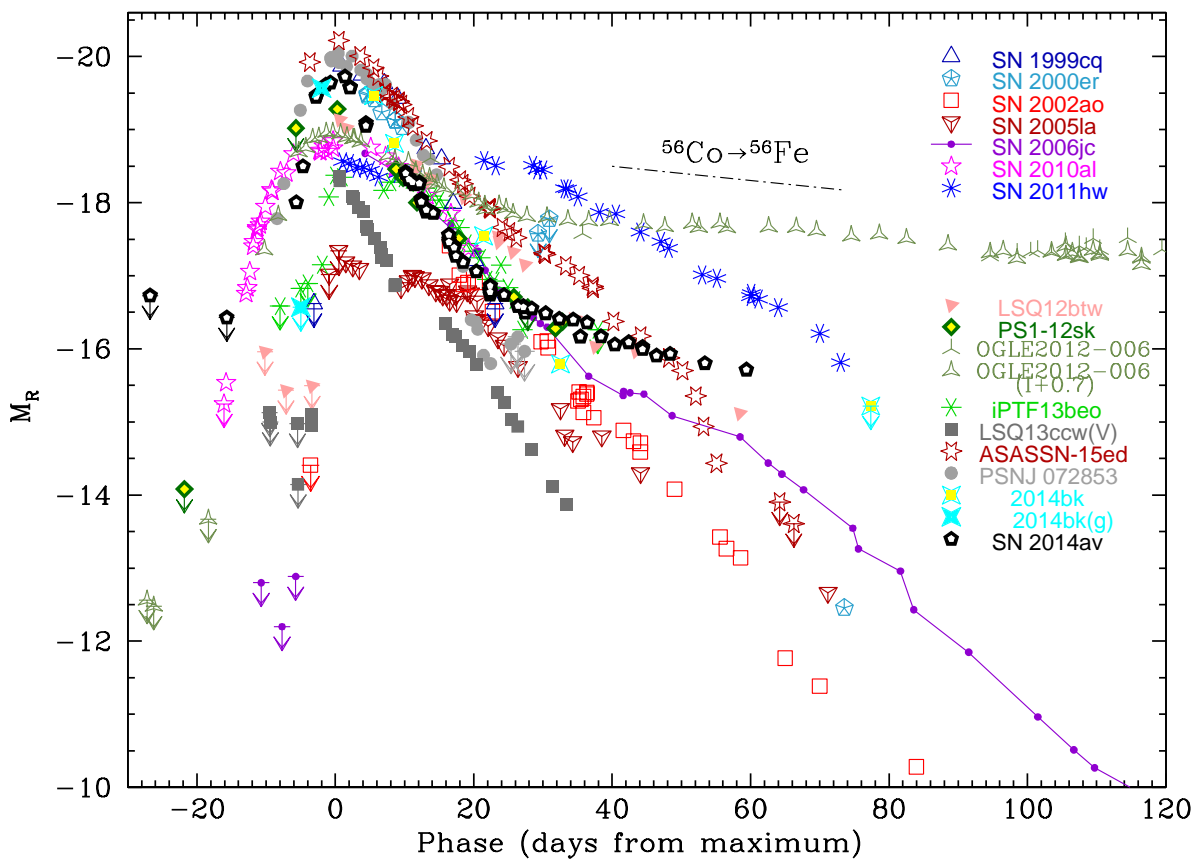}
\caption{$R$-band absolute light curves of SN~2014av and a wide sample of SNe Ibn (for the sources of the data, see notes underneath Table \ref{tab5}). 
The most significant pre-discovery limits are also shown. For OGLE-2012-SN-006, along with the
poorly sampled $R$-band light curve, the $I$-band light curve is also shown for completeness, scaled by $+$0.7 mag to approximately match the $R$-band points.
For LSQ13ccw, we showed the best sampled $V$-band light curve. SN~2014bk, whose data are shown here for the first time, is poorly sampled in the R band.
For this reason, we also report the closest detection limit of SN~2014bk and the $g$-band discovery magnitude announced in \protect\citet{moro14}.
\label{fig3}}
\end{figure*}

Thanks to the data collected by amateur astronomers, we were able to monitor the light curve rising phase, hence to provide an estimate of the explosion epoch.
Through a low-order polynomial fit, we found that the $R$-band maximum occurred on 2014 April 23.6 UT ($JD$ = 2456771.1 $\pm$ 1.2).
Fitting the earliest photometric data using a parabolic function, we estimated that the explosion occurred 10.6 days before
maximum, i.e. on  $JD$ = 2456760.5, which is fully consistent with the epoch suggested by the latest non-detection
on 2014 April 6.8 UT ($JD$ = 2456754.3).

\begin{table*}
\begin{center}
\caption{Main light-curve parameters for our sample of Type Ibn SNe, including absolute peak magnitudes (column 5), rise-time to the $R$-band maximum
(column 6), and decline rates at three different time intervals (columns 7 to 9).
Information on the reddening is provided in the reference papers; when not available, 
the Milky Way component is taken from \protect\citet{sch11}, while the host galaxy component is obtained by measuring the equivalent width
of the interstellar Na I doublet lines from the low resolution spectra available to us, adopting the low-reddening empirical relation
from \protect\citet{tur03}. The slopes are measured in units of mag 100$d^{-1}$.  
 \label{tab5} }
\begin{tabular}{cccccccccc}
\hline \hline
SN   & Type &  $\mu$ & $E(B-V)_{tot}$ & $M_{R,peak}$ & Rise time (d) & $\gamma^R_{0-25}$ & $\gamma^R_{25-60}$ & $\gamma^R_{60-150}$ & Sources \\ \hline
SN 1999cq         & Ibn     &  35.27 & 0.15 & $-$19.87  & $<$4  & 15.5$\pm$1.9 & --           & --           & 1 \\
SN~2000er         & Ibn     &  35.52 & 0.11 & $<-$19.49  & --    &  8.9$\pm$0.6 & --           & --           & 2 \\
SN~2002ao         & Ibn     &  31.73 & 0.25 & $<-$17.41  & --    & 12.6$\pm$2.5 & 9.8$\pm$0.3 & --           & 2 \\
SN~2005la         & Ibn/IIn &  34.49 & 0.01 & $-$17.19  & --    & non-monotonic & 7.5$\pm$0.6 & --           & 3 \\      
SN~2006jc         & Ibn     &  32.01 & 0.04 & $<-$18.61$^\ddag$  & $<$10 &  8.6$\pm$0.7 & 7.7$\pm$0.3 & 7.9$\pm$0.2&  3,4,5 \\
SN~2010al         & Ibn     &  34.27 & 0.06 & $-$18.86  & 16    &  9.4$\pm$0.6 & 12.6$\pm$0.2 & --           & 8 \\
SN~2011hw         & Ibn/IIn &  34.92 & 0.10 & $<-$18.54  & --    & non-monotonic & 5.5$\pm$0.1 & --           & 8 \\
PS1-12sk          & Ibn     &  36.84 & 0.03 & $-$19.21  & $<$23 & 10.8$\pm$0.9  & --          & --           & 9 \\ 
OGLE-006$^\star$   & Ibn     &  36.94 & 0.07 & $-$19.65 & 15.6    & 4.8$\pm$0.1 & 0.4$\pm$0.1 & 1.0$\pm$0.1   & 10 \\ 
LSQ12btw          & Ibn     &  36.97 & 0.02 & $-$19.14 & $<$4 & 7.3$\pm$0.4 & 5.8$\pm$0.7 & --               & 11 \\
iPTF13beo         & Ibn-pec &  38.01 & 0.04 & $-$18.39  &  --    & non-monotonic & 12.8$\pm$3.5  & --         & 12 \\
LSQ13ccw$^\odot$   & Ibn-pec &  37.07 & 0.04 & $-$18.36 & $<$6 & 12.6$\pm$0.2 & $\sim$15.5 & --               & 11 \\ 
CSS140421$^\dag$   & Ibn     &  37.41 & 0.03: & $-$19.4: &  --    & --    & --     & --                          & 13 \\
ASASSN-14dd$^\odot$& Ibn     &  34.23 & 0.15: & $-$19.1: & --    & --     &  --     & --                 & 14 \\ 
SN~2014av$^\ast$   & Ibn     &  35.56 & 0.02 & $-$19.75 & 10.6 & 12.1$\pm$0.7$^\ast$  & 3.0$\pm$0.2$^\ast$ & -- & 15 \\
SN~2014bk         & Ibn     &  37.40 & 0.05 & $<-$19.47 &  --    & 13.1$\pm$1.1  & --     & --               & 15 \\ 
ASASSN-15ed       & Ibn/Ib  &  36.59 & 0.14 & $-$20.19  & -- &  11.4 $\pm$ 0.2 &  7.7 $\pm$ 0.3  &  26.3 $\pm$ 2.5 & 16 \\
PSN J07285387+3349106 & Ibn &  33.85 & 1.02 & $-$19.95$^\diamondsuit$  & $>$8.7 &  19.7$\pm$1.6 &  --   & --               & 17 \\
SN~2015G          & Ibn/Ib  &  31.80 & 0.33: & $<-$17.1         & --    &  --  &  --        & --              & 18 \\
 \hline
\hline
\end{tabular}
\begin{flushleft}
$^\star$ OGLE-006 = OGLE-2012-SN-006 ($I$-band light curve data). $^\dag$ CSS140421 = CSS140421:142042+031602. $^\ddag$ From unpublished data kindly provided by K. Itagaki.
$^\odot$ For LSQ13ccw, we considered the $V$-band light curve information; for ASASSN-14dd, we considered the $V$-band discovery magnitude. $^\ast$ Average between the $R$- and $r$-band slopes.
$^\diamondsuit$ The uncertainty in the line-of-sight reddening to PSN J07285387+3349106 is extremely large, giving an error in the $R$-band absolute magnitude of $\pm$1.13 mag.\\
1 = \protect\cite{mat00};
2 = \protect\cite{pasto08a}; 
3 = \protect\cite{pasto08b}; 
4 = \protect\cite{pasto07};
5 =  \protect\cite{fol07};
8 = \protect\cite{pasto14b}; 
9 = \protect\cite{san13};
10 = \protect\cite{pasto14c};
11 = \protect\cite{pasto14a};
12 = \protect\cite{gor14}; 
13 = \protect\cite{pol14};
14 = \protect\cite{sta14};
15 = this paper;
16 = \protect\cite{pasto15d};
17 = \protect\cite{pasto15e};
18 = \protect\cite{yus15}. 
\end{flushleft}
\end{center}
\end{table*}

Using low-order polynomial fits, we also estimated the peak magnitude and the decline rate of SN~2014av in the different bands. 
As there is a clear change in the slope of the light curves of SN~2014av at $\sim$ 25 days after maximum, we
computed the SN magnitude decline rates in two time intervals: from the peak to 25 days after maximum, and from 25 days to the latest detections. 
The results are reported in Table \ref{tab4}.
We note, in particular, that the blue-band light curves decline faster than the red-band light curves.
This trend is more evident during the early decline ($\gamma_{0-25}$ in Table \ref{tab4}), while at later phases there is more homogeneity in the decline
rates. The decline rates in all bands during the temporal window 25-60 days ($\gamma_{25-60}$) are much faster than those
expected from the decay of $^{56}$Co into $^{56}$Fe. However, as we will discuss later in this paper,
other indicators suggest that during the entire monitoring period, the radioactive
decays are not the major powering source for the light curve of SN~2014av (see Section \ref{bolom}).

Additional near-infrared (NIR) photometry was obtained on 2014 May 16 (i.e. one month after the SN discovery) at the TNG, equipped with NICS.
The images were reduced following standard prescriptions, including flat-field correction and sky subtraction; individual dithered images for each band
were then combined to obtain a single, higher S/N frame.
The observations, calibrated using the 2MASS catalogue \citep{skr06}, provided the following photometric measurements:
$J$ = 18.33 $\pm$ 0.06, $J-H$ = 0.29 $\pm$ 0.07 and $J-K$ = 0.74 $\pm$ 0.13. We can note that, at this epoch, there is no signature
of a significant NIR flux excess.

\subsection{Absolute light curve} \label{ph_cfr_Ibn}

\begin{table*}
\begin{center}
\caption{Log of spectroscopic observations of SN~2014av. \label{tab6} }
\begin{tabular}{ccccccc}
\hline \hline
Obs. date   & Average JD & Days & Instrumental configuration & Exposure time & Range & Resolution \\
 & (+2400000) & (after maximum) & & (s) &  (\AA) & (\AA) \\
 \hline
2014/04/23 & 56771.14 & 0 & Lijiang 2.4m telescope + YFOSC + gm10 & 600 & 3560--9160 & 45 \\
2014/05/01 & 56779.04 & 7.9 & Lijiang 2.4m telescope + YFOSC + gm10 & 1200 & 3460--9250 & 45 \\
2014/05/03 & 56781.37 & 10.3 & 3.58m TNG + Dolores + LRB + LRR & 2400+1200 & 3250-10000 & 11;10 \\
2014/05/04 & 56782.11 & 11.0 & Lijiang 2.4m telescope + YFOSC + gm10 & 1238 & 4040--9270 & 45 \\
2014/05/10 & 56788.49 & 17.4 & 2.56m NOT + ALFOSC + gm4 & 2$\times$2700 & 3400--9090 & 18 \\
2014/05/19 & 56797.46 & 26.4 & 3.58m TNG + Dolores + LRR & 2700 & 5020--9740 & 14 \\
2014/05/27 & 56805.44 & 34.3 & 3.58m TNG + Dolores + LRB & 3600 & 3400--8060 & 11 \\ 
\hline
\end{tabular}
\end{center}
\end{table*}

\begin{figure*}
\includegraphics[width=13.5cm,angle=270]{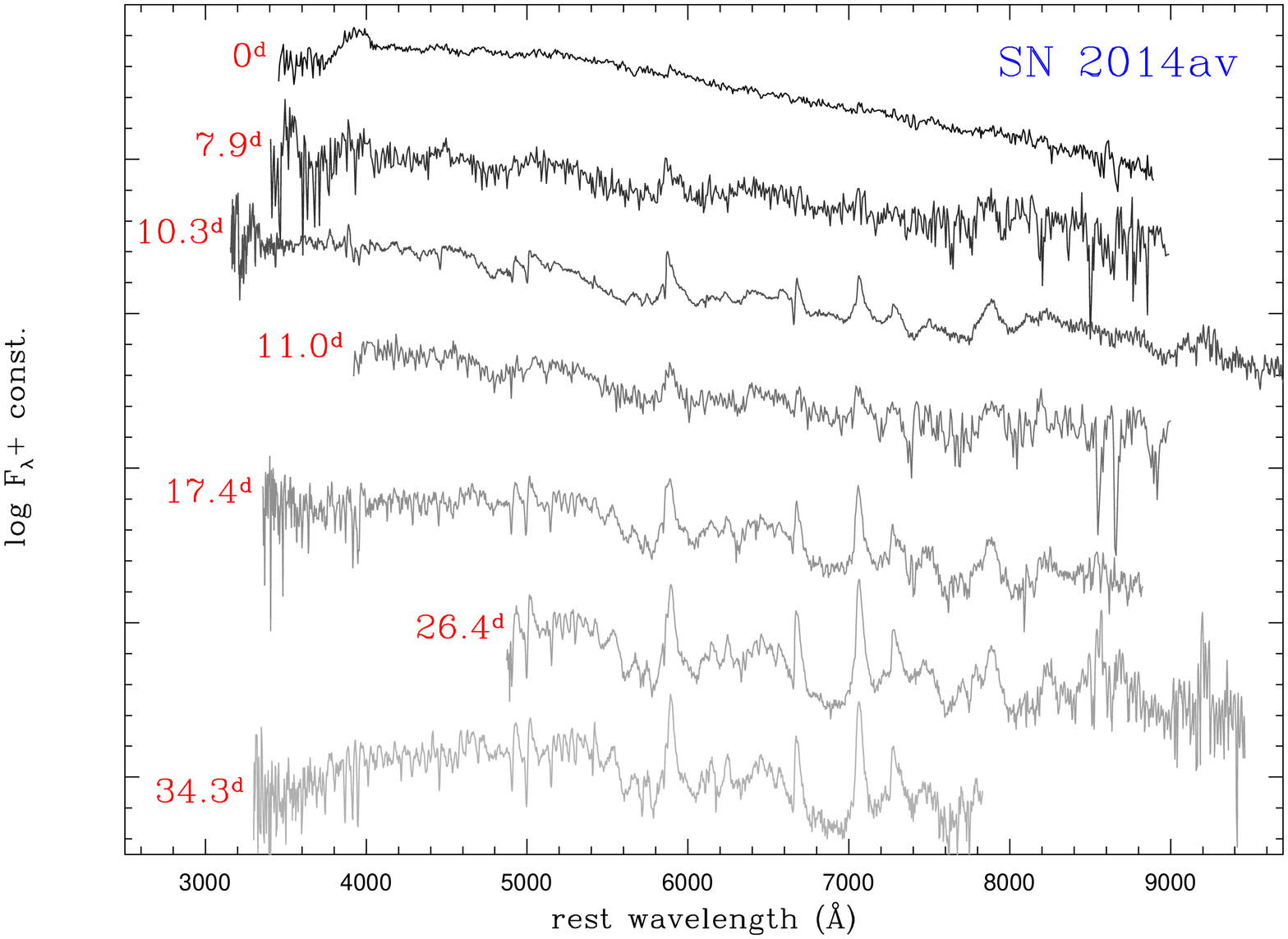}
\caption{Spectral evolution of SN~2014av. The spectra have been Doppler-corrected, although no reddening correction has been applied.
\label{fig4}}
\end{figure*}

SN~2014av was discovered soon after the explosion and a few days before maximum. 
Adopting the distance and reddening values reported in Section \ref{intro}, SN~2014av reached an absolute magnitude of $-19.76 \pm 0.16$ in the $R$ band, then
it experienced a  fast decline of $\sim$3 magnitudes. After about 25 days post-maximum, the light curve declined
with a slower  rate in all bands, as mentioned in Section \ref{ph} (see Table \ref{tab4}). A flattening of the optical light curve is unusual, but not unique in late
Type Ibn SNe, viz. in OGLE-2012-SN-006 \citep{pasto14c}. In order to better determine the photometric properties of SN~2014av in the context
of SNe Ibn, we measured the main parameters of the $R$-band light curves for a wide sample of objects. 
In particular, we estimated (when possible) 
the absolute peak magnitudes, the duration of the rise phase to maximum, and the post-peak declines at three temporal windows: between maximum and +25 days
($\gamma^R_{0-25}$), from +25 days and +2 months ($\gamma^R_{25-60}$), and at later phases ($\gamma^R_{60-150}$).
The results are reported in Table \ref{tab5}.  When $R$-band observations were incomplete to estimate the above parameters, we included measures obtained in
other bands, as specified in the footnote. From a rapid inspection of the table, we note that SNe Ibn are quite luminous, most of them 
having absolute $R$-band magnitudes close to $-$19, or exceeding that value.
Only one object, the transitional Type Ibn/IIn SN~2005la \citep{pasto08b}, appears to be significantly fainter than other SNe Ibn, although
for other objects we have incomplete information to derive firm conclusions on their peak luminosity.
However, despite most SNe Ibn are very luminous at maximum, the light curve shapes are widely heterogeneous.

\begin{figure*}
\includegraphics[width=17.6cm,angle=0]{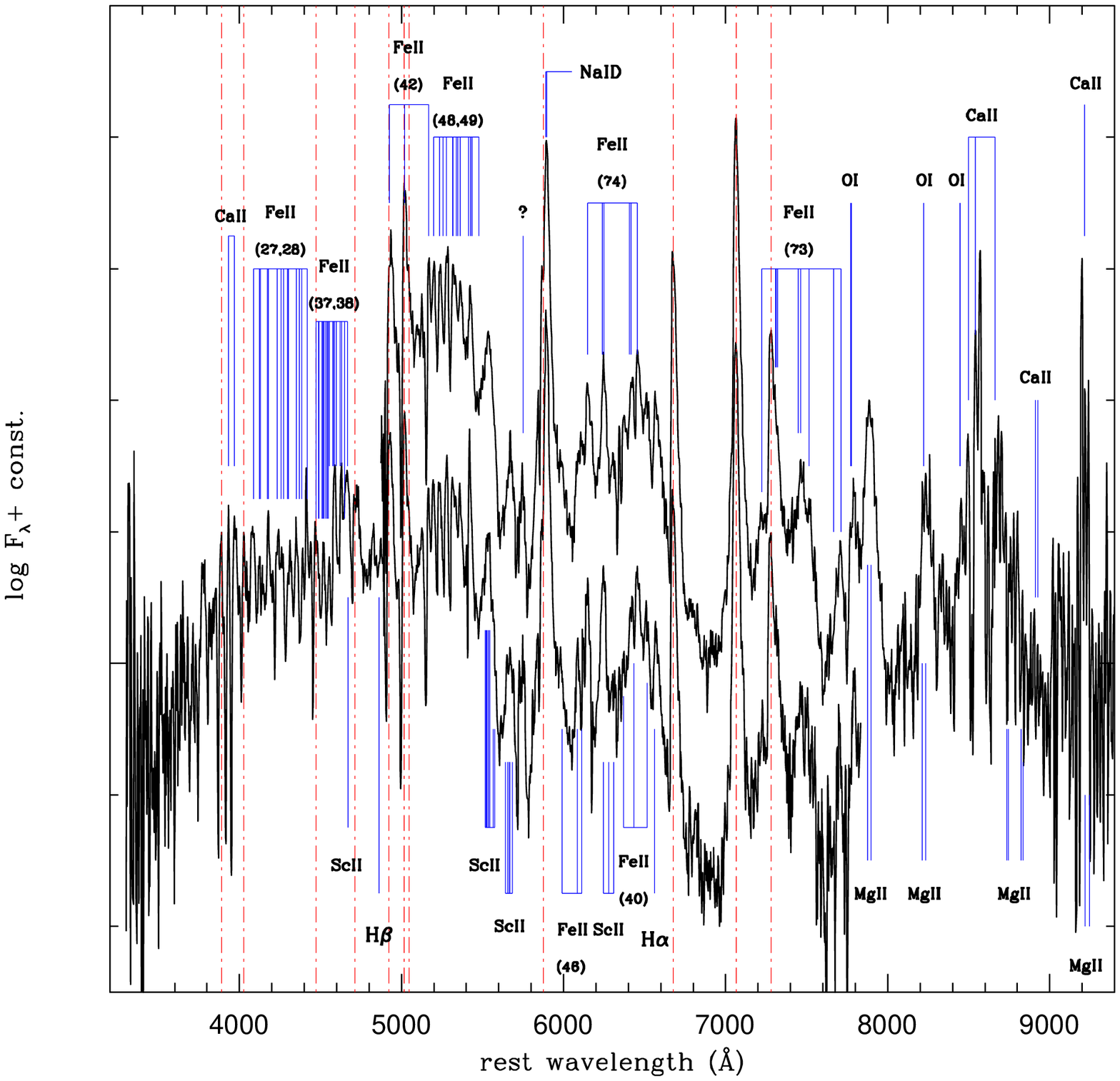}
\caption{Identification of the most prominent lines in the late-time TNG spectra of SN~2014av. Dot-dashed vertical lines mark the rest wavelength
positions of the strongest He I features.
\label{fig5}}
\end{figure*}

This can be best highlighted in Figure \ref{fig3}, where the absolute $R$-band light curve of SN~2014av is compared with those of a wide sample of SNe Ibn. We also show unpublished data of
the recent SN Ibn 2014bk \citep{moro14}, which is a relatively fast-evolving SN Ibn. Photometric data for this object are listed in Appendix \ref{app}. The comparison shows that
 the light curves are significantly different among the objects of this family. The absolute magnitudes at peak of these SNe mainly range 
from -18 \citep[iPTF13beo,][]{gor14} to -20 \citep[SN 1999cq,][]{mat00}.
Some objects (e.g. SN~2010al and OGLE-2012-SN-006) show a slow rise to maximum (up to 16 days), but other SNe Ibn experience a very fast rise to the peak ($\leq$6 days). SN~2014av has an intermediate 
rise time of about 10 days. The post-peak decline is even more heterogeneous, with one object having a very fast-declining light curve after maximum
\citep[LSQ13ccw,][]{pasto14a}, many with almost linear post-peak optical drops \footnote{In SN~2006jc  the increased late-time optical slope was interpreted as a signature 
of dust formation in a cool dense shell \citep{smi08,seppo08,elisa08}.}, and a few others showing double-phase light curve declines. The latter objects experience an initially fast decline, which 
is followed by a clear flattening at later stages. SN~2014av belong to this group, although the most extreme case is OGLE-2012-SN-006 \citep[][]{pasto14c}, with an early decline of about 5 mag/100$^d$, 
followed by a very long phase with a very flat light curve  (0.4 mag/100$^d$ from 25 and 60 days, and 1 mag/100$^d$ later on). 
As a comparison, the $R$-band decline rate of SN~2010al at phases $>$ 1 month after maximum is about 
3 mag/100$^d$, which is a factor 3 higher than the decline rate expected in a $^{56}$Co-powered event.
Finally, three objects (SN~2005la, SN~2011hw and iPTF13beo) showed a non-linear light curve decline after the first maximum, with at least one secondary luminosity peak 
\citep{pasto08b,smi12,gor14,pasto14b}.

\section{Spectroscopic Observations} \label{sp}

The spectroscopic monitoring campaign of SN 2014av started $\sim$4 days after the SN discovery, and continued for about one month. 
Spectra have been collected using the Lijiang 2.4m telescope of the  Yunnan Astronomical Observatory (YNAO) of the Chinese Academy of Sciences 
(equipped with YFOSC), the 3.58-m TNG with Dolores and the 2.56-m NOT with ALFOSC. Information on the spectroscopic observations is given in 
Table \ref{tab6}, the sequence of spectra available for SN~2014av is shown in Figure \ref{fig4}.

The earliest spectrum, obtained at maximum light (i.e. 10.6 days after the explosion), has poor S/N and modest resolution
(see Table \ref{tab6}). It is characterised by a blue continuum, with superposed narrow
and weak lines of He I, the most prominent being the $\lambda$5876 transition. Such line has a P-Cygni profile, whose minimum is 
blue-shifted by about $2100 \pm 800$~km~s$^{-1}$. A relatively strong feature detected at about 4700~\AA~(rest frame) is 
tentatively identified as He II $\lambda$4686. Another very prominent feature is detected at about 3950~\AA, and is tentatively identified as a blend of
He I and Ca II $\lambda\lambda$3934, 3968 (Ca II H$\&$K). 

Subsequent spectra (at phases 7.9 day to 11 day after maximum) show a remarkable evolution. A blue pseudo-continuum typical of Type Ibn
SNe and other interacting events is now visible. He I lines with low-contrast P-Cygni profiles are clearly detected: $\lambda$3889,
$\lambda$4026, $\lambda$4472, $\lambda$4713, $\lambda$4922,   $\lambda$5016 and $\lambda$5048. The He II $\lambda$4686 feature has now disappeared. Broader absorptions
are attributed to blends of Fe II (see also Figure \ref{fig5}).

At redder wavelengths, very prominent He I lines are detected, with the emission component which dominates over the P-Cygni absorption:
$\lambda$5876 (possibly blended with the Na~I doublet), $\lambda$6678, $\lambda$7065 and $\lambda$7281. The position of the deep minimum of the He I $\lambda$6678 feature
indicates a velocity of the He-rich wind of $940 \pm 110$~km~s$^{-1}$. However, the most prominent He I lines show a clear double-component
profile. From deblending the $\lambda$7065 line with 2 Gaussian components, we infer a broader component with full width at half maximum (FWHM) 
velocity v$_{FWHM} \approx$ 4500 km s$^{-1}$ (marginally evolving with time), with superimposed a narrow line with v$_{FWHM} \approx 1000$~km~s$^{-1}$ 
(not resolved in the YFOSC spectra).
Numerous additional bumps and individual lines are detected between 5500 and 7000 \AA, mostly due to Fe II lines. From these spectra
there is marginal evidence for the presence of a weak H$\alpha$. 

\begin{figure*}
\includegraphics[width=11.8cm,angle=270]{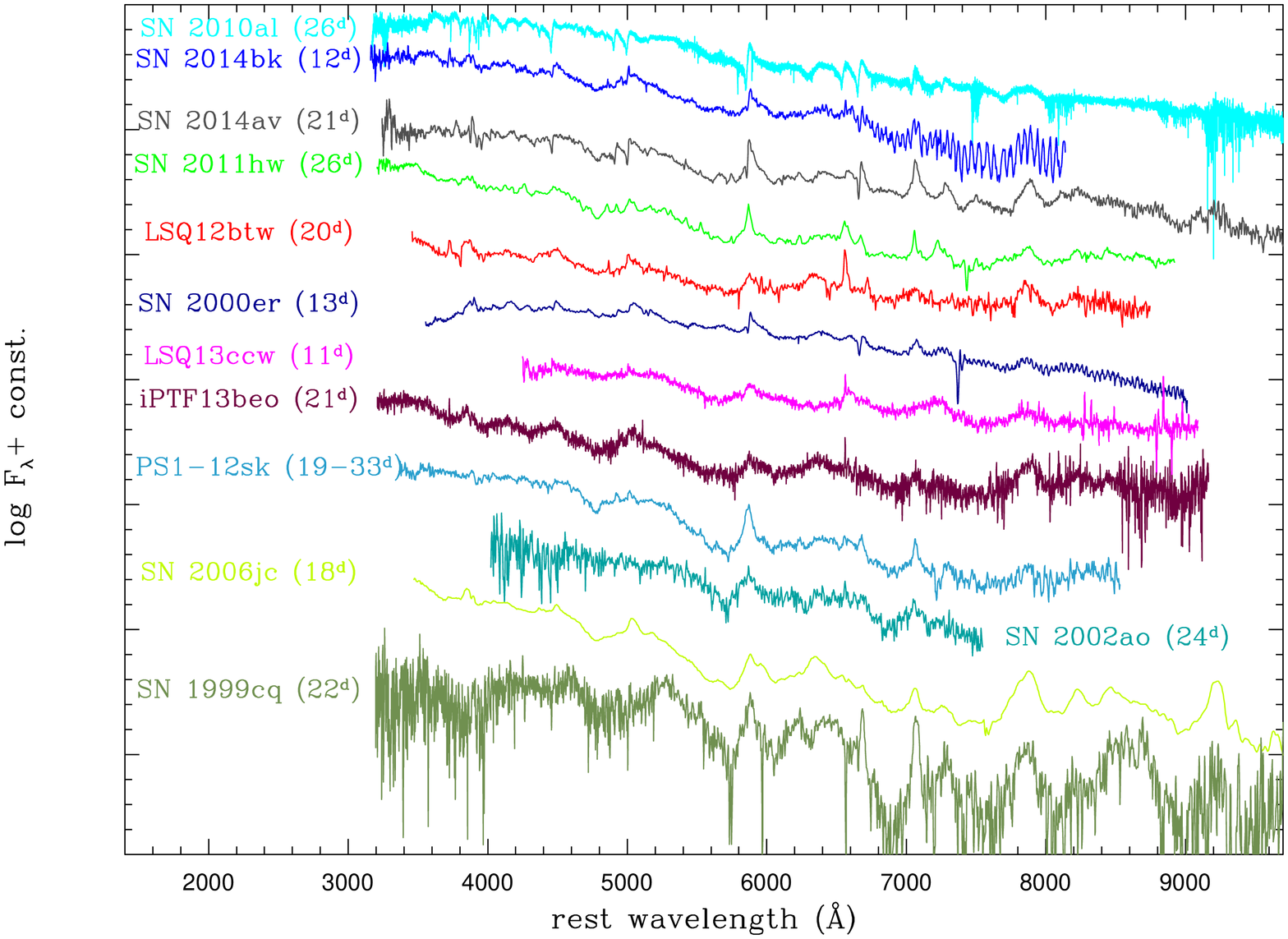}
\caption{Collection of spectra of Type Ibn SNe obtained at phases between 11 and 26 days: SNe 2014av and 2014bk (this paper), SNe 2010al and 2011hw \citep{pasto14b}, LSQ12btw and LSQ13ccw \citep{pasto14a},
SNe 2000er ans 2002ao \citep{pasto08a}, iPTF13beo \citep{gor14}, PS1-12sk \citep{san13}, SN~2006jc \citep{pasto07} and SN 1999cq \citep{mat00}.  
Phases reported in brackets are days from  the presumed explosion epochs. 
\label{fig6}}
\end{figure*}

At the reddest edge of the optical spectral domain, we notice a broad emission at about 7890 \AA, with v$_{FWHM} \approx$ 4700 km s$^{-1}$.
This is likely due to Mg II $\lambda$7877 and $\lambda$7896. Another shallow bump peaks at about 8200 \AA, which we identify as 
Mg II $\lambda$8214 and $\lambda$8235. The wide red tail of this feature is possibly due to other Mg II lines
($\lambda$8735, $\lambda$8746, $\lambda$8824 and $\lambda$8835), although we cannot rule out a contribution from Ca II $\lambda\lambda\lambda$8498,
8542, 8662 (hereafter the NIR Ca~II triplet). A further feature at about 9200 \AA~can be due to Mg II $\lambda$9218 and
Mg II $\lambda$9244.

At later epochs, the emissions become stronger, and allow us to perform a more robust line identification. Using our latest TNG spectra 
(phases 26.4 days and 34.3 days after maximum) we accurately identify the most prominent spectral features in Figure \ref{fig5}. We still see the prominent He I lines,
whose broad and narrow components have now  v$_{FWHM} \approx$ 2800 km s$^{-1}$ and v$_{FWHM} \approx$ 1200-1300 km s$^{-1}$, respectively.
Now Ca II H$\&$K and the NIR triplet are clearly discernible, with velocity of 1580 $\pm$ 230 km s$^{-1}$ (as measured from the positions
of the  minimum of the two Ca II H$\&$K lines).
Mg II lines are still quite prominent, but at these phases, also O I features are identified:  $\lambda\lambda\lambda$7772,7774,7775, $\lambda$8222,
$\lambda$8446 (partially blended with Ca II NIR). Fe II lines with P-Cygni profiles are still detected, with average velocities of  1230  $\pm$ 120 km s$^{-1}$.
From these later-epoch spectra, we also note the presence of Sc II lines. Again, we tentatively identify a weak H$\alpha$ emission, though 
alternative identifications cannot be ruled out (see also Section \ref{hydrogen}). In particular, assuming that it is H$\alpha$, its FWHM would be 820~km~s$^{-1}$, which is 
consistent with the velocity inferred for the Fe II lines, although marginally slower.

\subsection{Comparison with other Type Ibn SNe}

The  spectra of Type Ibn SNe are characterised by the presence of prominent and relatively narrow He I features.
The heterogeneity of the photometric properties of SNe Ibn has been remarked in Section \ref{ph_cfr_Ibn}. In Figure \ref{fig6}, we compare a collection of spectra of
SNe Ibn whose phases are approximately known, including an unpublished spectrum of SN~2014bk.\footnote{In analogy with other Type Ibn SNe, this spectrum is dominated
by He I lines mostly in emission, showing two components with different widths: a narrow component with a P-Cygni profile blue-shifted by 840 $\pm$ 140 km s$^{-1}$ is superposed on a much broader
component with v$_{FWHM}$ = 7860 $\pm$ 320 km s$^{-1}$.} The comparison in the figure highlights the existence of some heterogeneity among SNe Ibn also
in their spectral observables. First of all, the velocities of the most prominent line components range from about 700 km s$^{-1}$
to a few thousands km s$^{-1}$ (and these velocities may be significantly phase-dependent). The wide velocity range probably depends on the gas regions
where these lines originate, e.g. in the unperturbed CSM, in a shocked shell, in the shocked or unshocked SN ejecta or a combination of different emitting regions. 
We will further discuss the nature of the different components of the He I
lines in Section~\ref{disc}. In addition, there are clear differences in the strengths
of the broad $\alpha$-element lines, which are occasionally prominent \citep[e.g. in SN~2006jc,][]{pasto07,fol07}, and 
sometimes almost undetectable, like in the case of the Type Ibn/IIn SN~2005la \citep[][see Section \ref{alfa}]{pasto08b,mod14}.
In a few cases, narrow coronal lines were also detected in SNe Ibn \citep[for example in SN~2011hw,][]{smi12,pasto14b}.
Finally, there is heterogeneity in the spectroscopic evolution time scales among the objects of the sample,
which probably depends on the geometry and the distribution of the CSM with which the SN ejecta are interacting.

\section{Quasi-bolometric light curve and physical parameters} \label{bolom}

\begin{figure}
\includegraphics[width=8.6cm,angle=0]{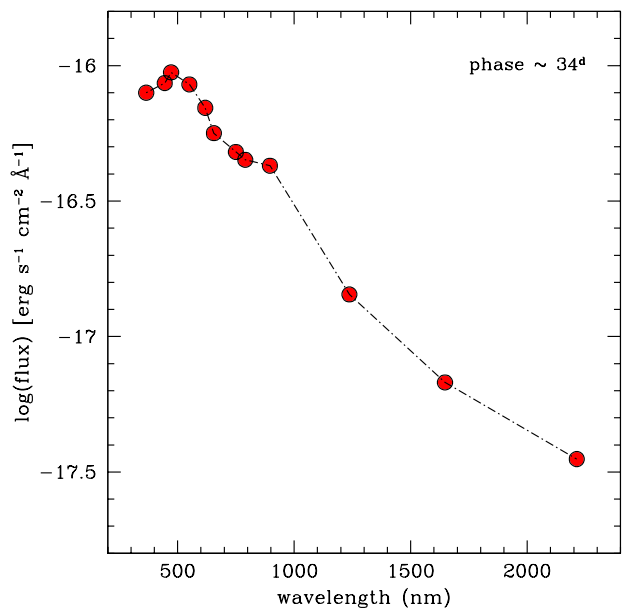}
\caption{SED of SN~2014av computed on May 16, 2014, i.e. about 34 days after the explosion.
\label{fig7}}
\end{figure}

In order to constrain some basic physical parameters for SN~2014av, we computed its pseudo-bolometric light curve by integrating the flux contribution 
of individual optical bands. For each band, we derived the flux at the effective wavelength, considering only epochs with Sloan $r$-band observations being
available. When individual photometric points at given epochs were not available, their contribution was computed  using the magnitude information in adjacent epochs. 
Since early-time photometry (during the rising branch and around the light curve peak) was not available in most photometric bands, the early flux contribution in the missing bands 
 was obtained assuming an early colour evolution of SN~2014av similar to that of the Type Ibn SN~2010al \citep{pasto14b}.

\begin{figure}
\includegraphics[width=8.74cm,angle=0]{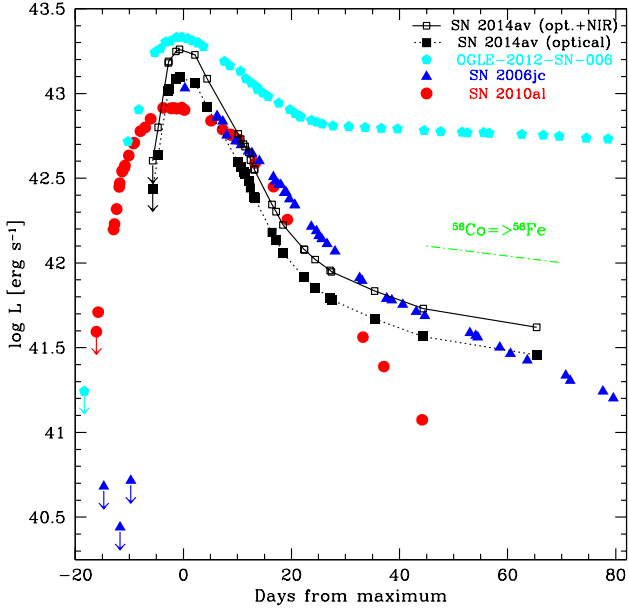}
\caption{Pseudo-bolometric light curve of SN~2014av, with (solid line) and without (dotted line) the additional NIR contribution, as specified in the text.
For comparison, also the bolometric light curves of the Type Ibn SNe 2006jc, 2010al and OGLE-2012-SN-006 are shown. For the comparison objects, the light curves
are computed by integrating the contributions from the $U$ to the $K$ bands. 
\label{fig8}}
\end{figure}

The fluxes provided the spectral energy distribution (SED) at the given phase. The observed fluxes were integrated with the trapezoidal rule 
and converted to luminosity adopting the distance and interstellar reddening estimated in Section \ref{intro}. 
We did not account for the UV flux contribution, although it might have been significant at early phases \citep[see, e.g.,][]{pasto14b},
while we did estimate the NIR flux using the single 2014 May 16 NIR observation, under the very rough assumption of constant NIR contribution.
The SED computed on May 16 is shown in Figure \ref{fig7}. The pseudo-bolometric light curve of SN~2014av is shown in Figure \ref{fig8} (with and without 
the NIR contribution), and is compared with the light curves of the Type Ibn SNe 2006jc \citep[data are from][]{pasto07,pasto08a,fol07,elisa08}, 2010al \citep{pasto14b}
and OGLE-2012-SN-006  \citep{pasto14c}. The most significant detection limits are also shown.

We confirm the large scatter in the photometric properties of SNe Ibn, as discussed in Sect. \ref{ph_cfr_Ibn}. Although SN~2014av (accounting from the NIR contribution) 
has a peak magnitude comparable with that of the very luminous OGLE-2012-SN-006, and is slightly more luminous at maximum than SNe 2006jc and 2010al, 
its photometric evolution is very fast. It experiences a very rapid rise to the light curve peak similar to that inferred for SN~2006jc, and
has a post-peak decline which is faster than any other SN in the sample. However, it clearly flattens at phases above 20-25 days, to a luminosity similar to that of SN~2006jc, 
but with a slope which is twice as fast as the $^{56}$Co to $^{56}$Fe decay rate. 

Assuming a modest contribution of the ejecta-CSM interaction to the observed luminosity evolution in SN~2014av, its high peak luminosity
($\sim 1.8 \times 10^{43}$ erg~s$^{-1}$) and its late-time light curve evolution  would be indicative of a moderate ejected $^{56}$Ni mass.
As there is a decent match with the light curve of SN~2006jc, we would obtain for SN~2014av an amount of  $^{56}$Ni
similar to the value inferred for SN~2006jc \citep[0.2-0.3 M$_\odot$,][]{pasto08a,tom08}.
This value for the $^{56}$Ni mass is relatively high for a core-collapse SN, although it has already been observed in some stripped-envelope SNe which ejected
several solar masses of material \citep[e.g. the ``hypernovae'',][]{nom06,pm13}. On the other hand, it is very unlikely that SN~2014av ejected a few solar masses of material.
In fact, it is well known that the evolutionary time-scale of the light curve in a non-interacting SN depends on the ejected mass to kinetic energy ratio \citep[e.g.][]{arn82}.
As a consequence, a narrow (i.e. fast-evolving) light curve peak would probably be indicative of modest ejected masses, with a significant fraction
of this material being composed by radioactive isotopes. 

We note, however, that large M($^{56}$Ni) / M$_{ej}$ ratios are typical of thermonuclear explosion rather than canonical core-collapse events, and
a thermonuclear explosion scenario is contradicted by other observables, including the preferential location of SNe Ibn in star-forming galaxies \citep[e.g.,][]{pasto14c},
the strengths of $\alpha$-element spectral features and the detection of a pre-SN eruption in SN~2006jc, which is supportive of a massive
progenitor. For all this,  Type Ibn SNe are most likely core-collapse events. However, since the energy released in the radioactive decays 
alone cannot comfortably explain  their observed evolution, alternative/additional powering mechanisms (including CSM-ejecta interaction) 
may provide more plausible explanations for the properties of SN~2014av and other Type Ibn SNe, as proposed by \citet{chu09}.

\section{Spectral characterization of Type Ibn supernovae} \label{sp_cha}

\subsection{He I line velocity evolution} \label{disc}

As discussed above and in other recent papers in the literature,
SNe Ibn display a very wide range of observed properties. 
This is not in contrast with expectations, as wide heterogeneity in the observable is also observed - for example - in Type IIn SNe. 
The physical parameters of SNe Ibn strongly depend on the geometry, the composition, and the density 
profile of the CSM, along with the mass and the composition of the residual stellar envelope 
at the time of the terminal SN explosion. 

A method to constrain the properties of the stellar wind and the nature of the line emitting regions
is studying the evolution of the velocity of the spectral lines. Spectra of SNe interacting with a 
CSM (such as Type IIn and Type Ibn events) show lines with multiple-width components. These are though
to be produced in different gas regions \citep{che85,che94,chu94,chu97}. Multiple components in the spectral lines, 
in fact, indicate that the emitting materials move at different velocities. 
Narrow lines (with velocities from a few hundreds to $\leq$ 2000 km s$^{-1}$) likely generate
in slowly expanding, photo-ionized material. Very likely, this gas is unshocked CSM produced by the progenitor star before exploding as a SN.  
More controversial is the location of intermediate to broad components (from a few thousands to $\sim$ 10$^4$  km s$^{-1}$):
they can either be produced in a gas interface between two shock fronts (forward and reverse shocks), or in the freely expanding SN ejecta.

In SNe Ibn, hence, the study of the line profiles provides insights on the the velocity of the emitting
material, and gives key information on the mass-loss history of the SN progenitors. When a clear P-Cygni profile 
is identified, the velocity of the  He-rich expanding material is obtained by measuring the position of the
core of the blue-shifted absorption. When this component is not detected, the velocity is estimated through the FWHM of 
the strongest He I emission lines, obtained after deblending the full line profile with Gaussian fits.

The evolution of the narrow components of the He I lines for the entire SN sample is shown in Figure \ref{fig9}, 
that of the intermediate-width components is shown in Figure \ref{fig10}. 
As velocities are measured from spectra available in the literature, in some cases only modest S/N spectra are
available. This explains the large error bars occasionally shown in the figures. The velocities of the narrow He I
component in SN~2011hw spectra are measured from the spectra published in \protect\citet{pasto14b}, with the exception of two
points in Figure \ref{fig9}, where we report measures performed by \citet{smi12} on moderate resolution spectra. 
For SN~2006jc, the spectra are taken from \citet{pasto07,pasto08a} and \citet{anu09}. 
A summary with the main outcomes from our inspection of the above SN spectra is reported in Table \ref{tab7}.

\begin{table*}
\begin{center}
\caption{Main properties of the spectra for the Type Ibn SN sample. In column 3, we report the velocities of the narrow He I line components,
in column 4 the range of velocities measured for the intermediate/broad He I components, while in column 5 we comment the robustness of 
the H$\alpha$ line detection associated with the SN.
 \label{tab7} }
\begin{tabular}{cccccc}
\hline \hline
SN   & Type &  v$_{narrow}$(HeI) (km s$^{-1}$) & v$_{broad}$(HeI) (range; km s$^{-1}$) & H$\alpha$ detection & source \\ \hline
SN 1999cq        & Ibn     & 1900 & 6150 & no  & 1 \\
SN~2000er        & Ibn     & 1000 & 3950-4300  & no & 2 \\
SN~2002ao        & Ibn     & 940  & 3500-6050 & weak & 2 \\
SN~2005la        & Ibn/IIn & 500  & 2000-4200 & strong & 3,a \\      
SN~2006jc        & Ibn     & 760  & 1700-3100 & weak & 3,4,5 \\
SN~2010al        & Ibn     & 1000-1250 & 2550-5800 & weak & 8 \\
SN~2011hw        & Ibn/IIn & 210-250  & 1350-2350 & moderate & 8 \\
PS1-12sk         & Ibn     & 130 & 3100-3300  & weak & 9 \\ 
OGLE-006         & Ibn     & 800-1000 & 2400-3250  & weak & 10 \\ 
LSQ12btw         & Ibn     & 970 & 3200-5250 & no & 11 \\
iPTF13beo        & Ibn-pec & 2070 & 4970  & no & 12 \\
LSQ13ccw         & Ibn-pec & 2300 & 6750 & uncertain & 11 \\ 
CSS140421        & Ibn     & unknown  & unknown & unknown & 13 \\  
ASASSN-14dd      & Ibn     & unknown  & unknown & unknown & 14 \\
SN~2014av        & Ibn     & 840-1240 & 4350-5000  & weak & 15 \\
SN~2014bk        & Ibn     & 1100 & 5950  & uncertain & 15 \\ 
ASASSN-15ed      & Ibn/Ib  & 1200-1500  & 6000-7000 & no & 16 \\
PSN J07285387+3349106 & Ibn & 1000-1400 & 3000-3450 & no & 17 \\
SN2015G          & Ibn/Ib  & $\sim$1300  & $\sim$5500 & no & 18,19 \\
 \hline
\hline
\end{tabular}

\begin{flushleft}
1 = \protect\cite{mat00};
2 = \protect\cite{pasto08a}; 
3 = \protect\cite{pasto08b}; 
a = \protect\cite{mod14};
4 = \protect\cite{pasto07};
5 = \protect\cite{anu09};
8 = \protect\cite{pasto14b}; 
9 = \protect\cite{san13};
10 = \protect\cite{pasto14c};
11 = \protect\cite{pasto14a};
12 = \protect\cite{gor14}; 
13 = \protect\cite{pol14}; 
14 = \protect\cite{pri14}; 
15 = this paper;
16 = \protect\cite{pasto15d};
17 = \protect\cite{pasto15e};
18 = \protect\cite{aya15}; 
19 = \protect\cite{fol15}. 

\end{flushleft}
\end{center}
\end{table*}

\subsubsection{Narrow He I line components}

\begin{figure}
\includegraphics[width=8.74cm,angle=0]{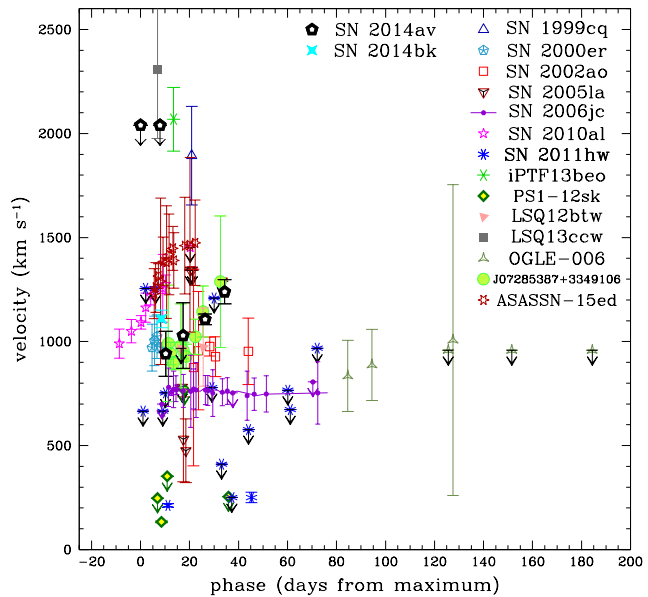}
\caption{Evolution of the velocity of the narrow He I line components for our SN sample. 
The reported velocities are computed as the weighted average of the values inferred for the individual 
He I lines. The errors are computed as the standard deviation of individual He I line velocities. 
In early spectra of Type Ibn SNe, the velocities for the narrow He I lines were preferentially estimated 
from the position of the absorption minimum. At later epochs, when narrow absorptions were no longer visible, we estimated 
their velocities from the FWHM of the emission component. 
\label{fig9}}
\end{figure}

The width of the narrow line components provides key information on the velocity of the unshocked He-rich CSM,
and hence allows us to directly probe the pre-SN stellar wind. The temporal evolution of the velocity of the narrow He I line
components is shown in Figure \ref{fig9}. When narrow components are not resolved in the spectra, we report in the figure
only the resolution limits. First of all, for most objects we note a very modest, if any, evolution with time. This is what we 
expect from an unshocked circumstellar emitting shell.

The most remarkable finding from Figure~\ref{fig9} (see also Table \ref{tab7}) is that the narrowest components 
observed in the spectra of our Type Ibn SN sample span a wide range of velocities. 
Objects showing low velocities for the unperturbed CSM are the two transitional type Ibn, viz. SNe 2011hw (200-250 km s$^{-1}$)
and 2005la (about 500~s$^{-1}$). In these two cases, an H$\alpha$ line associated with the SN is also detected, with moderate strength.
The identification of H lines along with modest stellar wind velocities (a few hundreds km s$^{-1}$) are compatible with a star which 
is transitioning between the LBV and the Wolf-Rayet (WNE-type) stages, as proposed by \citet{smi12}, \citet{pasto08b} and \citet{pasto14b}. 
Another object showing remarkably small velocities for the narrow CSM He I lines is PS1-12sk, from which a relatively weak intermediate-component 
H$\alpha$ (with v$_{FWHM} \sim$ 1600 km s$^{-1}$) is possibly detected in the post-peak spectra shown by \citet{san13} (see their figure 6).

\begin{figure}
\includegraphics[width=8.74cm,angle=0]{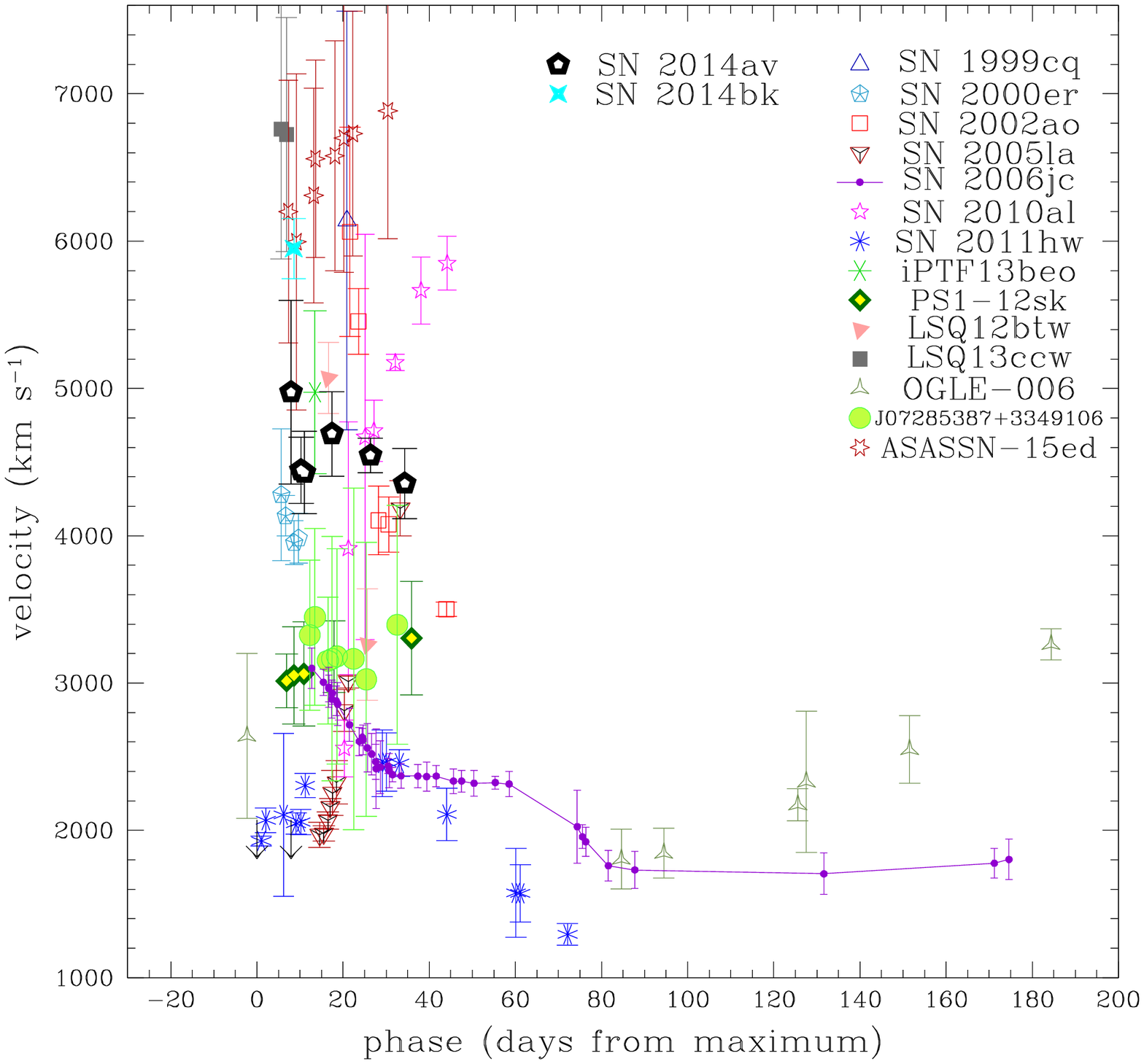}
\caption{Like Figure \ref{fig6}, but for the intermediate He~I line components. The reported velocities are 
computed as the weighted average among the measures available for the strongest He I lines. When a relatively broad P-Cygni 
absorption was clearly detected for the He~I lines, the velocity was estimated from the position of the minimum. When the broad 
absorption was not detected, the velocity was measured as the FWHM velocity of the intermediate/broad emission components
obtained after deblending the He~I line profile. Although we are aware that these non-homogeneous criteria adopted in measuring 
line velocities may undermine the robustness of the outcomes, this is the best we can do with the available data. With this
approach, we may offer for each SN an indicative estimate for the velocity of the fastest-moving detectable gas.
\label{fig10}}
\end{figure}

In most cases (including SN~2014av), the narrow He I components for our spectral sample have velocities around 800-1000 s$^{-1}$, while in three cases they widely
exceed that value. In fact, SN 1999cq, iPTF13beo and LSQ13ccw show narrow components with P-Cygni profiles with v$_{FWHM} \sim$ 1900-2300 km s$^{-1}$.
It is worth noting that two of these three objects have unusual properties, with iPTF13beo showing a double-peaked light curve \citep{gor14}, whilst LSQ13ccw 
has an extremely fast evolving light curve and some unique spectral features \cite[see discussion in][]{pasto14c} with some similarity with the 
peculiar Type Ib SN~2002bj \citep{poz10}. Type Ibn SNe, whose spectra show weak or no trace of H lines, and velocities inferred for the narrow He I
lines near or above 1000 km s$^{-1}$, most likely should be linked to more evolved Wolf-Rayet progenitors. 

\subsubsection{Intermediate/broad He I line components}

Figure \ref{fig10} shows the evolution of the expansion velocities as measured for the intermediate/broad components of the He I lines 
in our SN sample. The range of velocities for these components is reported in Table \ref{tab7}. They have velocities 
which are a factor  4-6 higher than those measured for the narrow components.

In contrast with the narrow features, the broader components of the He I lines are found to evolve 
significantly with time. In fact, the evolution of these components depends on the velocity of the ejecta and the density of
the interacting material. Their velocities may provide clues for the gas interface between two shock fronts, 
which also depends on the speed of the expanding SN ejecta.

In particular, it is interesting to inspect the evolution of the spectral lines of SN~2010al: at phases around the light curve peak, 
the velocity inferred from the mimimum of the He I lines is $\sim$ 1000 km s$^{-1}$ (Table \ref{tab7}), consistent with that
expected for the unshocked CSM. Thereafter, He I lines become broader with time,
increasing to $\sim$ 4700 km s$^{-1}$ at about 4 weeks after the light curve peak, and reaching 5800 km s$^{-1}$ at 1.5 months post-maximum.
Broad He I features with P-Cygni profile have been detected in ASASSN-15ed, with velocities of 6000--7000 km s$^{-1}$ (as inferred from the position of
the core of the P-Cygni absorptions). As the SN is a transitional event from Type Ib to Type Ibn,
these broad features are likely indicative of the speed of the expanding SN ejecta \citep{pasto15d}.
An increasing velocity of the intermediate/broad He~I components is observed in the Type Ibn/IIn SN~2005la, 
with the gas velocity rising from about 2000 km s$^{-1}$ soon after discovery to $\sim$4200 km s$^{-1}$ about 3 weeks later. 

However, in some cases, the He I lines become narrower with time (e.g., in SN~2002ao the He I intermediate component
decreases its width by a factor 2 in 3 weeks). 
This is also observed in SN~2014av, although its intermediate-velocity component has a more modest velocity evolution,
spanning from  5000 to 4300 km s$^{-1}$ in about 4 weeks. A similar moderate evolution of the He I intermediate components 
towards lower velocity values has been observed in SN~2006jc (from 3100 to 1700 km s$^{-1}$ in about 4 months).
All of this suggests a decline in the velocity of the shocked gas regions, which may result from an increased density profile
of the CSM gas. More complex is the evolution of the intermediate-component velocity of SN~2011hw. It  grows from 1900 to 2500 km s$^{-1}$
during the first month, but then it declines to about 1300 km s$^{-1}$ at 70 days from the discovery.
Again, this peculiar evolution can be ascribed to a complex density profile of the shocked gas region.

\subsection{Detection of H$\alpha$} \label{hydrogen}

The detection of H$\alpha$ (and, eventually, other Balmer lines) is crucial to accurately classify interacting transients.
Other ingredients for the classification are the detection of He I lines, and the relative strengths between H and He I features.
Although narrow He I emission lines are frequently detected in the spectra of Type IIn SNe\footnote{Narrow He I have been observed 
in the spectra of a number of SNe IIn covering a wide range of observables, including - for example - SNe 1994W, 1995G, 1995N, 1998S, 2005ip, 2006jd, 2007rt, 2009kn and 2010jl
\citep{fas01,fra02,pasto02,chu04,tru09,smi09,str12,kan12,zha12}.}, the relative strengths between
 H$\alpha$ and the most prominent optical He I lines (in particular $\lambda$5876 and $\lambda$7065) determines the classification of the transient as a Type IIn,
a Type Ibn or even a transitional object between these two SN types.

Balmer H lines are clearly detected in the two transitional Type Ibn/IIn SNe 2005la and 2011hw \citep{pasto08b,pasto14b,smi12}. 
In the spectra of these objects, the strength of H$\alpha$ is comparable with
those of the most prominent He I lines. The velocity evolution inferred for the H lines in SN~2005la is very similar to that of the He I lines. In the early spectra 
of SN~2005la published by \citet{mod14}, H$\alpha$ showed a narrow, marginally resolved (v$_{FWHM} \sim$ 400 km s$^{-1}$) component, superposed on a broader
base with a P-Cygni profile. The velocities deduced for the broader component evolved significantly with time, from about 1400  km s$^{-1}$ in the spectra
close to the explosion epoch, to about 4000  km s$^{-1}$ in the spectra obtained about 3 weeks later.
The highest resolution spectrum of SN~2011hw \citep{pasto14b} showed an unresolved H$\alpha$ (v$_{FWHM} <$ 250 km s$^{-1}$) atop of broader wings that followed
the velocity evolution observed for the He I lines (shown in Figure \ref{fig10}): the velocity of the intermediate H$\alpha$ component had a non 
monotonic evolution in the range between 1350 and 2350 km s$^{-1}$.

However, other Type Ibn events show no prominent H SN lines or, in some cases, the identification of H$\alpha$ is controversial,
with an alternative identification of the feature as C II being plausible. In particular, blends of H$\alpha$ with C II $\lambda$6580 cannot be ruled out 
in most SNe of our sample, including SN~2014av. In particular, the detection of multiple lines of C II in the spectra of PS1-12sk \citep{san13}
argued - at least in the case of that object - against the identification of H lines. When the discrimination between H$\alpha$ and 
C II $\lambda$6580 is problematic, the identification of H$\alpha$ is indicated as ``uncertain'' in Table \ref{tab7}.

\subsection{Detection of $\alpha$-element features} \label{alfa}

The robust detection of lines from heavier elements, in particular $\alpha$-elements such as O I, Mg II, Ca II and occasionally even Si II, has been confirmed
for a number of Type Ibn SNe. In most cases, features from these ions have been observed in emission, with FWHM velocities higher or similar to those of the
intermediate-width components of the He I lines (in the range between 2000 and 6000 km s$^{-1}$, depending on the SN and its phase).

Relatively broad O~I and Mg~II lines are frequently detected as strong emission features in the spectra of SNe Ibn. 
In particular, in SNe 1999cq and 2006jc, O~I and Mg~II lines are very prominent and with velocities  v$_{FWHM} \approx$ 5000-9000 km s$^{-1}$ 
\citep[see, e.g.,][]{mat00,pasto07}. 
These lines are clearly detected also in SN~2014av, though they are narrower than in other SNe Ibn, and with Mg II
lines being stronger than O I features. A high relative strength of Mg II vs. O I lines has also been observed in the reddened
PSN J07285387+3349106 \citep{pasto15e}.
We note, however, that O~I and Mg~II lines are weak in other SNe Ibn, including SN~2011hw, PS1-12sk, LSQ13ccw and LSQ12btw.
Finally, in a few cases (e.g. in SNe 2000er and 2005la), the non-detection of these lines is ascribed to
the early phases of the available spectra.

Also, the detection of the Ca II NIR feature is likely phase dependent, since this triplet is normally observed 
in spectra obtained a few weeks after the explosion. For this reason, the Ca II NIR remains undetected (or is very weak) 
in several objects, including SN~2000er, iPTF13beo, LSQ12btw, LSQ13ccw and PS1-12sk. 

Less ubiquitous is the identification of Si II in the spectra of Type Ibn SNe. While Si II lines are clearly detected
in a few objects (e.g. SNe 1999cq, 2000er, 2006jc LSQ12btw, iPTF13beo), these  are not detected in other SNe
of this class, including PS1-12sk, LSQ13ccw, SN~2005la and OGLE-2012-SN-006.

\section{About the environments and Type Ibn SN progenitors} \label{prog}

\citet{pasto14a} made a preliminary attempt to characterise the host galaxies of the sample of SNe Ibn considered in this paper.
All SNe Ibn have  been discovered in spiral galaxies, with the remarkable exception of PS1-12sk, which exploded in 
the outskirts of an elliptical galaxy \citep{san13}.\footnote{Although this locations would favour the association of PS1-12sk with old stellar population, 
we cannot rule out the association with a very low surface luminosity dwarf galaxy companion.} Accounting for the great predominance of
spiral galaxies among the hosts of SNe Ibn, it is natural to associate this SN type with massive stellar population.

SN 2014av exploded in a spiral (Sb-type) galaxy, the most luminous one ever observed hosting a SN Ibn \citep[$M_B = -21.8$,][]{pasto14a}.
The oxygen abundance inferred at the SN position is about 9.2 \citep{pasto14a}, suggesting a metal-rich environment. However, other metallicity
estimates of Type Ibn SN environments led to different conclusions (see, e.g., Appendix \ref{app2}).
\citet{pasto14a} estimated an average metallicity at the SN position of $<$~12~+~log~(O/H)~$>$~=~8.63~$\pm$~0.42 for the sample,
which is not particularly sub-solar, although there is a wide range in the values inferred for individual SNe. Using a smaller 
sample, \citet{tad15} inferred a slightly lower oxygen abundance, viz. $<$~12~+~log~(O/H)~$>$~=~8.45~$\pm$~0.10.
A large dispersion in the values of the oxygen abundance suggests that metallicity plays a marginal role in producing SNe Ibn. 

The detection of a major outburst before the explosion of SN~2006jc \citep{pasto07,fol07} was one of the arguments used to
support the association of the SN with an erupting massive Wolf-Rayet star. In other words, the precursor of SN~2006jc was 
very likely a Wolf-Rayet with a residual LBV-like behaviour. Although direct evidence of similar pre-SN eruptions
{\bf is} still missing in other Type Ibn SNe, the study of the properties of the He-rich CSM presented in this paper (including 
the constraints on the line velocities and the secure identification of $\alpha$-elements) supports massive
progenitors for most SNe Ibn. However, the observed differences in their photometric and spectroscopic behaviour suggest 
some heterogeneity in the properties of the progenitor stars. In particular, the evidence of variable signatures of H in 
the CSM, the fact that the CSM velocities inferred from the narrow He I lines span one order of magnitude (from 200 to 2000
km s$^{-1}$, see Figure \ref{fig9}) and the variable amount of He still present in the stellar envelope at the moment of the SN explosion
indicate that a wide range of sub-types of Wolf-Rayet stars, spanning from the transitional Opfe/WN9 stars to more stripped WC/O types 
\citep[see][and references therein]{smi12}, can very likely produce Type SNe Ibn SNe. In this context, 
observations of the transitional WN to WC-type Wolf-Rayet star NaSt1 \citep[also known as Wolf-Rayet 122,][]{mau15} suggest 
that the binary interaction may favour the transition among Wolf-Rayet sub-types via pre-SN bursts, favouring the heterogeneity in 
the final Wolf-Rayet properties soon before their core-collapse.  Finally, the similarity of - at least - SN Ibn LSQ13ccw with the peculiar
SN Ib 2002bj \citep[for which a helium detonation on a white dwarf scenario has been proposed by][]{poz10} has been mentioned in \citet{pasto14a}.

\section{Summary and final remarks} \label{summary}

We have presented optical spectroscopic and photometric data of the recent, well-monitored Type Ibn SN~2014av.
The object was discovered a few days before the light-curve maximum, and monitored for 2 months after the peak. Deep pre-explosion non-detection
limits, along with a few photometric points obtained during the rising phase to the maximum light, allowed us to estimate the explosion 
epoch with a very small uncertainty to be JD = 2456760.5. 

SN~2014av is one of the most luminous SNe Ibn in our sample, reaching an absolute magnitude $M_R = -19.78$ and a bolometric luminosity of about 1.8 $\times$ 10$^{43}$ erg s$^{-1}$. 
Despite the remarkable intrinsic luminosity, it is a relatively fast-evolving transient. In fact, although the rise time to maximum is not extremely short (10.6 days), 
the post-peak decline is very fast (see Tables \ref{tab4} and \ref{tab6}).
The spectra of SN~2014av evolve from showing an almost featureless continuum at early phases, to being dominated by intermediate-velocity He I lines, 
with relatively broad features from $\alpha$-elements. These later spectra are characterized by the unusual presence of a multiplicity of Fe II lines showing P-Cygni profiles.

SN~2014av has been compared with other objects classified as Type Ibn events, and - not unexpectedly - we have observed that a large heterogeneity exists among the objects of 
this class (in analogy with that observed in Type IIn SNe). In particular, 
\begin{itemize}
\item although most SNe Ibn are luminous ($M_R << -18$), there are rare exceptions of significantly weaker events (such as SN 2005la);
\item differences are observed in the photometric evolutionary time-scales, with objects showing extremely fast-evolving light curves \citep[such as that of LSQ13ccw;][]{pasto14a}
and others with very slow-evolving light curves resembling those of SNe IIn \citep[e.g., OGLE-2012-SN-006;][]{pasto14c};
\item light curves of SNe Ibn may show non-monotonic post-peak declines \citep[see, e.g., the  double-peaked light curve of iPTF13beo;][]{gor14}; 
\item a range of FWHM velocities is observed for the narrow He I line components of SNe Ibn, indicating intrinsic differences in the progenitor wind velocities;
\item relatively broad He I components can be detected in Type Ibn SN spectra, suggesting the presence of residual He in the
envelope of their progenitor stars;
\item in a few cases, SNe Ibn spectra have circumstellar Balmer lines, suggesting the presence
of a small and variable fraction of H in the composition of their CSM \citep{smi08,pasto08b,pasto14b}.
\end{itemize}
The heterogeneous observed parameters of Type Ibn SNe likely depend on the different pre-explosion configuration and chemical composition of their progenitor systems. 

The host of SN~2014av is a luminous ($M_B = -21.8$) Sb-type galaxy. In most cases, the galaxies hosting SNe Ibn are spirals, hence environments with active star formation, 
with the remarkable exception of PS1-12sk \citep{san13} that  exploded in the outskirts of an elliptical galaxy.
For this reason, we favour the association of SN~2014av and other Type Ibn SNe with a massive stellar population. Thus, Wolf-Rayet stars of different sub-types
are natural candidates to be the precursors of Type Ibn SNe.
As discussed in \citet{pasto14a}, the oxygen abundance at the SN position is super-solar, being about 9.2, the highest
in the Type Ibn SN sample. Since most Type Ibn SNe had exploded in environments showing a broad metallicity range (7.8 $<$ 12+log(O/H) $<$ 9.2),  metallicity
has very likely a marginal role in producing SNe Ibn.

\section*{Acknowledgements}

We acknowledge Gianpiero Locatelli, Stan Howerton, William Wiethoff, Jean-Marie Llapasset 
({\it http://www.astrosurf.com/jmllapasset/index.htm}),
Gianluca Masi, Francesca Nocentini and Patrick Schmeer (Virtual Telescope Project facility, at the Bellatrix Astronomical Observatory, see
websites {\it http://www.virtualtelescope.eu/} and {\it http://www.bellatrixobservatory.org/})
for kindly providing us their observations of SN~2014av. We also thank Mr. Toru Yusa for his help in collecting amateur astronomer images.

AP, SB, NER, AH, LT, GT, and MT are partially supported by the PRIN-INAF 2014 with the project “Transient Universe: unveiling 
new types of stellar explosions with PESSTO”. XW is supported by the Major State Basic Research Development Program (2013CB834903),
the National Natural Science Foundation of China (NSFC grants 11073013, 11178003, 11325313), and the Foundation of Tsinghua University
(2011Z02170). NER acknowledges the support from the European Union Seventh Framework Programme (FP7/2007-2013) under grant agreement n. 267251 "Astronomy Fellowships in Italy" (AstroFIt). JJZ is supported by the National Natural Science Foundation of China (NSFC, grant 11403096).

 This paper is based on observations made with the Italian Telescopio Nazionale Galileo
(TNG) operated on the island of La Palma by the Fundaci\'on Galileo Galilei of
the INAF (Istituto Nazionale di Astrofisica). It is also based on observations made with
the Nordic Optical Telescope (NOT), operated on the island of La Palma jointly by Denmark, Finland, Iceland,
Norway, and Sweden, in the Spanish Observatorio del Roque de los Muchachos of the Instituto de Astrof\'isica de Canarias;
the 1.82m Copernico Telescope of INAF-Asiago Observatory; and
on observations made with the Gran Telescopio Canarias (GTC), installed in the Spanish Observatorio del Roque de los Muchachos 
of the Instituto de Astrof\'isica de Canarias, in the Island of La Palma.
The Liverpool Telescope is operated on the island of La Palma by Liverpool John Moores University in the
Spanish Observatorio del Roque de los Muchachos of the Instituto de Astrofisica
de Canarias with financial support from the UK Science and Technology Facilities
Council.
We acknowledge the support of the staff of the Li-Jiang 2.4-m telescope (LJT). 
Funding for the LJT has been provided by Chinese academe of science (CAS) and the People's Government of Yunnan Province.  

We acknowledge the usage of the HyperLeda database (http://leda.univ-lyon1.fr).
This publication makes use of data products from the Two Micron All Sky Survey, which is a joint project of the University of Massach
usetts and the Infrared Processing and Analysis Center/California Institute of Technology,
funded by the National Aeronautics and Space Administration and the National Science Foundation.

\appendix

\section{Optical photometry of SN 2014bk} \label{app}
In Table \ref{tabapp}, we report  sparse photometry available for the Type Ibn SN 2014bk, included
in the SN sample discussed in this paper. These data have been reduced following standard
prescriptions, as described in Section \ref{ph}.

\begin{table*}
\begin{center}
\caption{Optical Johnson-Cousins magnitudes of SN~2014bk, and associated errors. \label{tabapp} }
\begin{tabular}{cccccccc}
\hline \hline
Obs. date   & Average $JD$ & $U$ & $B$ & $V$ & $R$ & $I$ & Instrument \\
 \hline
2014/06/05  &    2456813.55 &      --      &      --      &       --      &  17.976 (0.236)     &      --      &  1 \\ 
2014/06/08  &    2456816.54 & 18.021 (0.076) & 18.782 (0.054) & 18.635 (0.047) & 18.631 (0.077) & 18.274 (0.071) & 1 \\
2014/06/20  &    2456829.43 & 19.591 (0.087) & 20.096 (0.130) & 19.729 (0.075) & 19.898 (0.203) & 19.685 (0.234) & 1 \\
2014/07/02  &    2456840.55 & 21.240 (0.185) & 21.485 (0.144) & 21.226 (0.126) & 21.649 (0.282) & 20.381 (0.268) & 1 \\
2014/08/15  &    2456885.37 &      --      &      --      &       --      &   $>$22.22  &  -- & 2 \\ \hline
\end{tabular}
\\
\begin{flushleft}
1 = 2.56-m NOT + ALFOSC (La Palma, Canary Islands, Spain);
2 = 10.4-m GTC + OSIRIS (Sloan $r$-band observation transformed into Johnson-Cousins data; La Palma, Canary Islands, Spain). 
\end{flushleft}
\end{center}
\end{table*}

\section{Spectrum of the galaxy hosting SN 2014bk} \label{app2}
\begin{figure*}
\includegraphics[width=13.0cm,angle=0]{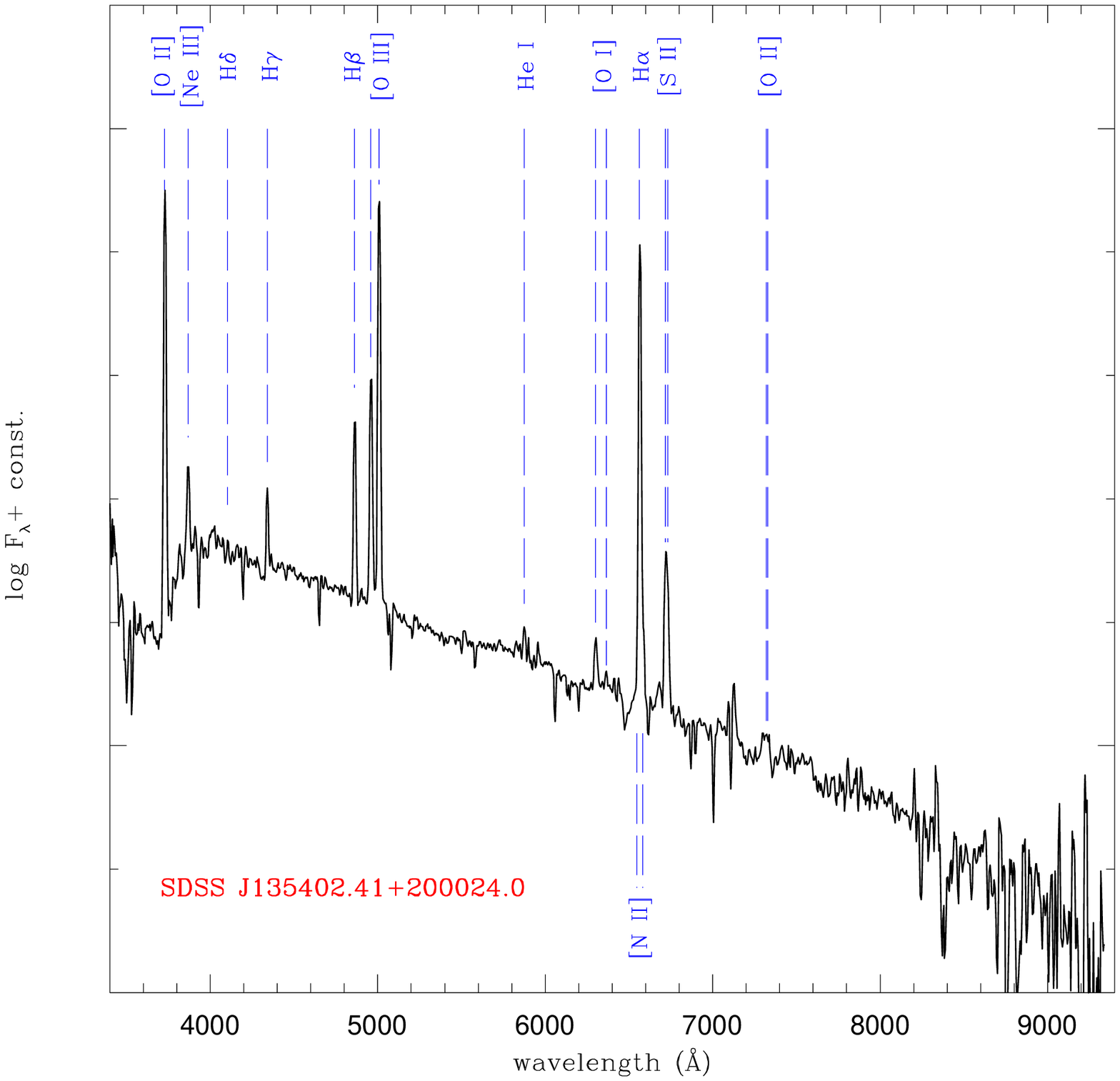}
\caption{GTC (Osiris) spectrum of the galaxy hosting SN 2014bk, centered at the location of the SN.
\label{figA1}}
\end{figure*}

A late-time spectrum of the comparison Type Ibn SN 2014bk was obtained at the 10.4-m Gran Telescopio Canarias (GTC) of
the Observatorio del Roque de los Muchachos in La Palma (Canary Islands). The spectrum (shown in Figure \ref{figA1}) was obtained on 2014 August 15 
(JD = 2456885.39) using OSIRIS and the R300B grism (resolution 17 \AA); it shows no signature of the SN, but is useful to
constrain the oxygen abundance at the SN location. The host galaxy, SDSS J135402.41+200024.0, has an estimated redshift z = 0.0697
\citep{moro14}.

In the spectrum, we securely identified several emission lines, the strongest being [O II] $\lambda$3927, [O III] $\lambda$5007 and H$\alpha$. 
Line fluxes for the strongest lines are reported in Table \ref{meta}. In order to estimate  the oxygen abundance,
we measured the fluxes of a few relevant lines (including H$\alpha$, H$\beta$ [O III] $\lambda$5007 and [N II] $\lambda$6584), 
and applied the {\it N2} and {\it O3N2} methods desctibed in \citet{pet04}.
The two indicators provide very consistent estimates, resulting in an average aboundance $<$~12~+~log~(O/H)~$>$~=~8.11 dex, which is largely
sub-solar. 

\begin{table*}
\begin{center}
\caption{Observed fluxes for the strongest emission lines in the GTC + Osiris spectrum of SDSS J135402.41+200024.0. \label{meta}}
\begin{tabular}{ccc} \hline \hline
Line   & $\lambda$ (\AA) & Flux (10$^{-16}$ erg s$^{-1}$ cm$^{-1}$)\\ \hline
$[$O II$]$ & 3727 & 18.56 $\pm$ 0.54 \\
$[$Ne III$]$ & 3868 & 2.41 $\pm$ 0.34 \\
H$\gamma$ & 4340 & 1.70 $\pm$ 0.23 \\
H$\beta$ & 4861 & 5.01  $\pm$ 0.26 \\
$[$O III$]$ & 4959 & 6.55 $\pm$ 0.42 \\
$[$O III$]$ & 5007 & 17. 89 $\pm$ 0.57 \\
H$\alpha$ & 6563 & 16.37 $\pm$ 0.90 \\
$[$N II$]$ &  6584 & 0.69 $\pm$ 0.32 \\
$[$S II$]$ & 6717,6731 & 4.26 $\pm$ 0.31 \\
$[$O II$]$ & 8320,8330 & 0.78 $\pm$ 0.25 \\ 
\hline
\end{tabular}
\\
\end{center}
\end{table*}

\label{lastpage}

\end{document}